\documentclass[]{revtex4}
\usepackage{graphicx}
\usepackage{fancyhdr}
\begin{document}
\def\ds{\displaystyle}
\def\beq{\begin{equation}}
\def\eeq{\end{equation}}
\def\bea{\begin{eqnarray}}
\def\eea{\end{eqnarray}}
\def\beeq{\begin{eqnarray}}
\def\eeeq{\end{eqnarray}}
\def\ve{\vert}
\def\vel{\left|}
\def\ver{\right|}
\def\nnb{\nonumber}
\def\ga{\left(}
\def\dr{\right)}
\def\aga{\left\{}
\def\adr{\right\}}
\def\lla{\left<}
\def\rra{\right>}
\def\rar{\rightarrow}
\def\nnb{\nonumber}
\def\la{\langle}
\def\ra{\rangle}
\def\bdll{$B_c \rightarrow D_{q'}^*l^+ l^-$}
\def\ba{\begin{array}}
\def\ea{\end{array}}
\title{ {\bf ANALYSIS OF THE RARE  $B_c \rightarrow
D_{s,d}^{*}~l^+ l^-$ DECAYS IN QCD   }}
\author{ K. Azizi  \\ Department of  Physics, Middle
East Technical University, 06531 Ankara, Turkey
(e146342@metu.edu.tr)\\F.
Falahati \\
 Physics Department,
Shiraz University,  Shiraz  71454,  Iran (phy1g832889
@shiraz.ac.ir)\\V. Bashiry
\\ Cyprus International University, Via Mersin 10 , Turkey (bashiry@ciu.edu.tr)\\S.M. Zebarjad\\
 {\small\sl Max-Planck-Institut f\"{u}r Physik  } (Werner
Heisenberg Institut), {\small\sl
F\"ohringer Ring 6}, {\small\sl 80805 M\"unchen, Germany}\\
\small permanent address: Physics Department, Shiraz University,
Shiraz 71454, Iran (zebarjad @physics.susc.ac.ir)}

\setlength{\baselineskip}{24pt} \maketitle
\setlength{\baselineskip}{7mm}

 The rare $B_c \rightarrow
D_{s,d}^{*}~l^+ l^-$ decays are investigated in the framework of the
three point QCD sum rules approach. Considering the gluon condensate
corrections to the correlation function, the form factors relevant
to these transitions are calculated.  The total decay width and
branching ratio for these decays  are also evaluated. The results
for the branching ratios are in good agreement with the quark
models.

\section{Introduction}
The rare $B_c \rightarrow D_{s,d}^{*}~l^+ l^-$ decays are
proceeded by flavor changing neutral current (FCNC) transitions of
$b\rightarrow s,d$. In the standard model (SM), these transitions
occur at loop level and are not allowed in the tree level. This
provides the most crucial framework to test the SM \cite{Buchalla,
Ali}. Among the $B$ mesons, the $B_{c}$ decays have received great
attention for the following reasons:

a) This meson constitutes a very rich laboratory for studying
various decay channels, which are essential from both theoretical
and experimental aspects. At LHC when it begins  operation with the
luminosity values of ${\cal L}=10^{34}cm^{-2}s^{-1}$ and
$\sqrt{s}=14\rm TeV$, the number of $B_c^{\pm}$ mesons is expected
to be about $10^{8}\sim10^{10}$ per year \cite{Du, Stone}, so there
are basic facilities to study not only some rare $B_c$ decays, but
also CP violation, T violation and polarization asymmetries.

b) $B_{c}$ decays could be used for a determination of the
Cabibbo-Kobayashi-Maskawa (CKM) matrix elements $V_{tq}$ (q = d,
s, b).

c) It is the lowest bound state of two heavy quarks (b and c) with
open  flavor. This is also the reason why $B_{c}$ decays weakly and
not strongly or electromagnetically.

d) It attracts the interest of physicists for checking predictions
of the pertubative QCD in the laboratory.

e) The $b\rightarrow s,d$ transitions are very sensitive to the
physics beyond the SM since some new particles might have
contributions in the loops diagrams.

 Some possible
channels of $B_{c}$ decays are $B_{c}\rightarrow l
\overline{\nu}\gamma$, $B_{c}\rightarrow \rho^{+}\gamma$,
$B_{c}\rightarrow K^{\ast+}\gamma$, $B_{c}\rightarrow
B_{u}^{\ast}l^{+}l^{-}$, $B_{c}\rightarrow B_{u}^{\ast}\gamma $ and
$B_{c}\rightarrow D_{s,d}^{\ast}\gamma $ which have been studied in
the framework of light-cone  and three point QCD sum rules
\cite{Aliev1, Aliev2, Aliev3, Alievsp, azizi1}. Larger sets of
exclusive non--leptonic and semi--leptonic decays of the $B_{c}$
meson have been studied within a relativistic constituent quark
model in Ref-~\cite{Ivanov}. This study describes the rare $B_c
\rightarrow D_{s,d}^{*}~l^+ l^-$ decays  in the framework of the
three point QCD sum rules approach. Here,  $l=e,~\mu,~\tau$ and
$D_{s,d}^{*}$ are vector mesons. Analyzing  these transitions could
give valuable information about the nature of the vector $D_{s}^{*}$
meson.

In $B_c \rightarrow D_{s,d}^{*}~l^+ l^-$ decays, the long distance
dynamics are parameterized by transition form factors, calculation
of which is a central problem for these decays. To calculate the
form factors, we use the three point QCD sum rules approach (for
details about this method see \cite{Shifman1, Colangelo}). This
method has been successfully  applied for various problems
\cite{braun1, aliev4, aliev5, Ozpineci, aliev6, aliev7, aliev8,
aliev9}. Note that, these transitions have been analyzed in the
framework of the relativistic constituent quark models (RCQM)
\cite{tt1}, light front quark model (LFQM)and constituent quark
model (CQM) \cite{tt2}.

The paper encompasses three sections: In section II, we calculate
the expressions for transition form factors, where the light quark
condensates don't contribute but the gluon correction
contributions are considered. The numerical analysis, discussion
and comparison of our results with the predictions of other quark
models are presented in section III.
\begin{figure}
\vspace*{-1cm}
\begin{center}
\includegraphics[width=13cm]{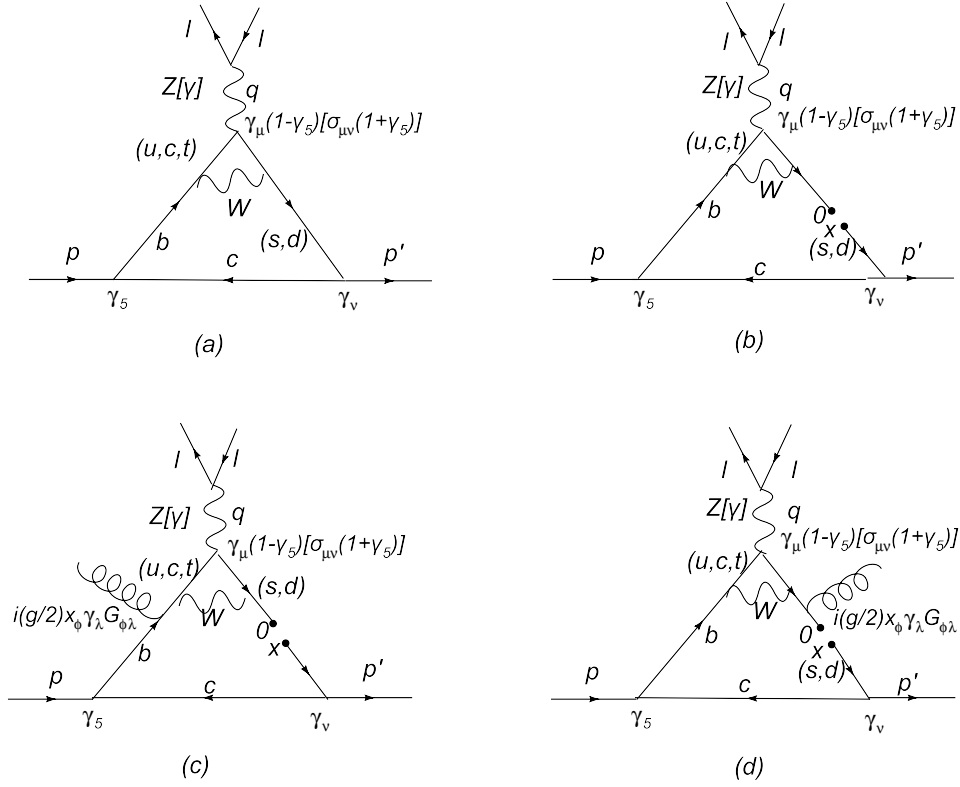}
\end{center}
\caption{Loop diagrams for $B_c \rightarrow D_{s,d}^{*}~l^+ l^-$
transitions, bare -loop (diagram a) and light quark condensates
(without any gluon diagram b and with one gluon emission diagrams c,
d)} \label{fig1}
\end{figure}

\section{Sum rules for the  \bdll transition form factors }
In the SM, the effective Hamiltonian for $B_c \rightarrow
D_{q'}^{*}l^+ l^-$$(q'=s,d)$  decays which occur via $b \rightarrow
q' l^+ l^-$
 loop  transition   can be written as:
\begin{equation}
H_{eff}= \frac{G_{F}\alpha}{2\sqrt{2}\pi} V_{tb}V_{tq'}^{*}\Bigg[
C_9^{eff}  \overline {q'} \gamma_\mu (1-\gamma_5) b~  \overline {l
}\gamma_\mu l + C_{10} \overline {q'} \gamma_\mu (1-\gamma_5) b~
\overline {l } \gamma_\mu \gamma_5 \l - 2 C_7^{eff}\frac{m_b}{q^2}
\overline {q'} ~i\sigma_{\mu\nu}
q^\nu (1+\gamma_5) b~  \overline {l}  \gamma_\mu l \Bigg],\\
\end{equation}
where $C_7^{eff}$, $C_9^{eff}$ and $C_{10}$   are Wilson
coefficients
 related to the $Z$ and photon penguin (see Fig. 1) and box diagrams. For more
about the Wilson coefficients see \cite{ Aliev3, tt1} and references
therein.
 The transition amplitude  of the $B_c \rightarrow D_{q'}^{*}l^+ l^-$ is
  obtained by sandwiching of the effective Hamiltonian between the initial and final states
\begin{eqnarray}\label{falahat}
\nonumber M &=& \frac{G_{F}\alpha}{2\sqrt{2}\pi}
V_{tb}V_{tq'}^{*}\Bigg[ C_9^{eff}
<D_{q'}^{*}(p',\varepsilon)\mid\overline{q'} \gamma_\mu (1-\gamma_5)
b\mid B_{c}(p)>  \overline {l }\gamma_\mu l + C_{10}
<D_{q'}^{*}(p',\varepsilon)\mid\overline {q'} \gamma_\mu
(1-\gamma_5) b\mid B_{c}(p)> \overline {l } \gamma_\mu \gamma_5 l \\
&-& 2 C_7^{eff}\frac{m_b}{q^2} <D_{q'}^{*}(p',\varepsilon)\mid
\overline {q'}  ~i \sigma_{\mu\nu} q^\nu (1+\gamma_5) b\mid
B_{c}(p)> \overline {l} \gamma_\mu l \Bigg],
\end{eqnarray}
where $p$ and $p'$ are the initial and final meson states,
respectively, and $\varepsilon$ is the polarization vector of
$D_{q'}^{*}$ meson. Our aim is to calculate the matrix elements
appearing in Eq.~(\ref{falahat}). From Lorentz invariance and
parity conservation point of view, these matrix elements can be
parameterized in terms of the form factors in the following way:
\begin{eqnarray}\label{3au}
\nonumber <D_{q'}^{*}(p',\varepsilon)\mid\overline
{q'}\gamma_{\mu}(1-\gamma_{5}) b\mid
B_c(p)>&=&\frac{2A_{V}(q^2)}{(m_{B_{c}}+m_{D_{q'}^{*}})}\varepsilon_{\mu\nu\alpha\beta}
\varepsilon^{\ast\nu}p^\alpha p'^\beta-iA_{0}(q^2)(m_{B_{c}}
+m_{D_{q'}^{*}})\varepsilon_{\mu}^{\ast} \\
+i\frac{A_{+}(q^2)}{(m_{B_{c}}+m_{D_{q'}^{*}})}(\varepsilon^{*}p)P_{\mu}
&+&i\frac{A_{-}(q^2)}{(m_{B_{c}}+m_{D_{q'}^{*}})}(\varepsilon^{*}p)q_{\mu},
\end{eqnarray}
\begin{eqnarray}\label{4au}
\nonumber <D_{q'}^{*}(p',\varepsilon)\mid\overline
{q'}\sigma_{\mu\nu} q^\nu (1+\gamma_5) b\mid
B_{c}(p)>&=&2T_{1}(q^2)~i\varepsilon_{\mu\nu\alpha\beta}
\varepsilon^{\ast\nu}p^\alpha p'^\beta+T_{2}(q^2)\Bigg\{\varepsilon_{\mu}^{\ast}(m_{B_{c}}^2-m_{D_{q'}^{*}}^2)\\
-(\varepsilon^{*}p)P_{\mu}\Bigg\}
&+&T_{3}(q^2)~(\varepsilon^{*}p)\Bigg\{q_{\mu}-\frac{q^2}{m_{B_{c}}^2-m_{D_{q'}^{*}}^2}P_{\mu}\Bigg\},
\end{eqnarray}
where $A_{V}(q^2)$, $A_{0}(q^2)$, $A_{+}(q^2)$ , $A_{-}(q^2)$,
$T_{1}(q^2)$, $T_{2}(q^2)$ and $T_{3}(q^2)$ are the transition form
factors. $P_{\mu}=(p+p')_{\mu}$ and $q_{\mu}=(p-p')_{\mu}$, here,
$q$ is the momentum of the $Z$ boson (photon). Also, for simplicity,
we redefine those as:
\begin{eqnarray}\label{eq5}
A'_{V}(q^2)&=&\frac{2A_{V}(q^2)}{(m_{B_{c}}+m_{D_{q'}^{*}})}~,~~~~~~~~~~~~A'_{0}(q^2)=A_{0}(q^2)(m_{B_{c}}
+m_{D_{q'}^{*}}),
\nonumber\\
A'_{+}(q^2)&=&-\frac{A_{+}(q^2)}{(m_{B_{c}}+m_{D_{q'}^{*}})}~,~~~~~~~~~
A'_{-}(q^2)=-\frac{A_{-
}(q^2)}{(m_{B_{c}}+m_{D_{q'}^{*}})},\nonumber\\
T_{V}'(q^2)&=&-2T_{1}(q^2)~,~~~~~~~~~~~~~~~~~~~T_{0}'(q^2)=-~T_{2}(q^2)(m_{B_{c}}^2-m_{D_{q'}^{*}}^2),\nonumber\\
T_{-}'(q^2)&=&-~T_{3}(q^2).
\end{eqnarray}
For the calculation of these  form factors,  the QCD sum rules
method is applied. Following the general philosophy of the QCD sum
rules, we start by considering the following correlators:
\begin{eqnarray}\label{6au}
\Pi _{\nu\mu}^{V-A}(p^2,p'^2,q^2)&=&i^2\int
d^{4}xd^4ye^{-ipx}e^{ip'y}<0\mid T[J _{\nu D_{q'}^{*}}(y)
J_{\mu}^{V-A}(0) J_{B_{c}}(x)]\mid  0>,\nonumber\\ \Pi
_{\nu\mu}^{T-PT}(p^2,p'^2,q^2)&=&i^2\int
d^{4}xd^4ye^{-ipx}e^{ip'y}<0\mid T[J _{\nu D_{q'}^{*}}(y)
J_{\mu}^{T-PT}(0) J_{B_{c}}(x)]\mid 0>,
\end{eqnarray}
where $J _{\nu D_{q'}^{*}}(y)=\overline{c}\gamma_{\nu} q'$ and
$J_{B_{c}}(x)=\overline{b}\gamma_{5}c$   are interpolating
currents of the initial and final meson states, respectively.
 $J_{\mu}^{V-A}=~\overline {q'}\gamma_{\mu}(1-\gamma_{5})b $ ~and~
 $J_{\mu}^{T-PT}=~\overline {q'}{\sigma_{\mu\nu} q^\nu (1+\gamma_5)}b
 $ are the vector, axial vector, tensor and pseudo tensor parts of the transition currents.
To calculate the phenomenological part of the correlators given in
Eq.~(\ref{6au}), two complete sets of intermediate states with the
same quantum numbers as the currents $J_{D_{q'}^{*}}$ and
$J_{B_{c}}$ are inserted. As a result of this procedure, we get
the following representation of the above-mentioned correlators:
\begin{eqnarray} \label{7au}
&&\Pi _{\nu\mu}^{V-A}(p^2,p'^2,q^2)=-
\nonumber \\
&&\frac{<0\mid J_{D_{q'}^{*}}^{\nu} \mid
D_{q'}^{*}(p',{\varepsilon})><D_{q'}^{*}(p',{\varepsilon})\mid
J_{\mu}^{V-A}\mid B_{c}(p)><B_{c}(p)\mid J_{Bc}\mid
0>}{(p'^2-m_{D_{q'}^{*}}^2)(p^2-m_{Bc}^2)}+\cdots,
\nonumber \\
&&\Pi _{\nu\mu}^{T-PT}(p^2,p'^2,q^2)=-
\nonumber \\
&&\frac{<0\mid J_{D_{q'}^{*}}^{\nu} \mid
D_{q'}^{*}(p',{\varepsilon})><D_{q'}^{*}(p',{\varepsilon})\mid
J_{\mu}^{T-PT}\mid B_{c}(p)><B_{c}(p)\mid J_{Bc}\mid
0>}{(p'^2-m_{D_{q'}^{*}}^2)(p^2-m_{Bc}^2)}+\cdots,
\end{eqnarray}
where $\cdots$ represents contributions coming from higher states
and continuum. The matrix
 elements $<0\mid J_{D_{q'}^{*}}^{\nu} \mid
D_{q'}^{*}(p',{\varepsilon})>$ and $<B_{c}(p)\mid J_{Bc}\mid 0>$ are
defined  in the standard way as:
\begin{equation}\label{8au}
 <0\mid J^{\nu}_{D_{q'}^{*}} \mid
D_{q'}^{*}(p',{\varepsilon})>=f_{D_{q'}^{*}}m_{D_{q'}^{*}}\varepsilon^{\nu}~,~~<B_{c}(p)\mid
J_{Bc}\mid 0>=-i\frac{f_{B_{c}}m_{B_{c}}^2}{m_{b}+m_{c}},
\end{equation}
where $f_{D_{q'}^{*}}$ and $f_{B_{c}}$  are the leptonic decay
constants of $D_{q'}^{*} $ and $B_{c}$ mesons, respectively. Using
Eq. (\ref{3au}), Eq. (\ref{4au}) and Eq. (\ref{8au}) and performing
summation over the polarization of the $D_{q'}^{*}$ meson, for the
phenomenological (physical) part of the correlation function we
obtain
\begin{eqnarray}\label{9amplitude}
\Pi_{\nu\mu}^{V-A}(p^2,p'^2,q^2)&=&-\frac{f_{B_{c}}m_{B_{c}}^2}{(m_{b}+m_{c})}\frac{f_{D_{q'}^{*}}m_{D_{q'}^{*}}}
{(p'^2-m_{D_{q'}^{*}}^2)(p^2-m_{Bc}^2)} \times
\left[\vphantom{\int_0^{x_2}}A'_{0}(q^2)g_{\mu\nu}+A'_{+}(q^2)P_{\mu}p_{\nu}\right.
\nonumber
\\ &+&\left. A'_{-}(q^2)q_{\mu}p_{\nu}+i\varepsilon_{\mu\nu\alpha\beta}p^{\alpha}p'^{\beta}A'_{V}(q^2)\vphantom{\int_0^{x_2}}\right] + \mbox{excited
states,}\nonumber\\
\Pi_{\nu\mu}^{T-PT}(p^2,p'^2,q^2)&=&-\frac{f_{B_{c}}m_{B_{c}}^2}{(m_{b}+m_{c})}\frac{f_{D_{q'}^{*}}m_{D_{q'}^{*}}}
{(p'^2-m_{D_{q'}^{*}}^2)(p^2-m_{Bc}^2)} \times
\left[\vphantom{\int_0^{x_2}}-i~T'_{0}(q^2)g_{\mu\nu}\right.
\nonumber\\
&-&
\left.i~T'_{-}(q^2)q_{\mu}p_{\nu}+\varepsilon_{\mu\nu\alpha\beta}p^{\alpha}p'^{\beta}T'_{V}(q^2)\vphantom{\int_0^{x_2}}\right]
+ \mbox{excited states.}
\end{eqnarray}
Next, we calculate the correlation function in the quark and gluon
languages via the operator product expansion (OPE) which is called
the QCD or theoretical side of the correlator. For this reason the
correlation function is written as:
\begin{eqnarray}\label{QCD side}
\Pi_{\nu\mu}^{V-A}(p^2,p'^2,q^2)&=&\Pi^{V-A}_{0}g_{\mu\nu}+\Pi^{V-A}_{+}P_{\mu}p_{\nu}+
\Pi^{V-A}_{-}q_{\mu}p_{\nu}+i\Pi^{V-A}_{V}\varepsilon_{\mu\nu\alpha\beta}p^{\alpha}p'^{\beta},
\nonumber\\
\Pi_{\nu\mu}^{T-PT}(p^2,p'^2,q^2)&=&-i~\Pi^{T-PT}_{0}g_{\mu\nu}-i~\Pi^{T-PT}_{-}q_{\mu}p_{\nu}+\Pi^{T-PT}_{V}
\varepsilon_{\mu\nu\alpha\beta}p^{\alpha}p'^{\beta},
\end{eqnarray}
where each $\Pi_{i}$ with $i=0, \pm$ and $V$ is  defined in terms of
the perturbative and nonpertubative parts as follows
\begin{eqnarray}\label{QCD side1}
\Pi_{i}&=&\Pi_{i}^{pert}+ \Pi_{i}^{nonpert}.
\end{eqnarray}
For calculating the perturbative part of the correlator, we consider
the bare -loop diagram (Fig. 1 (a)), as for the nonperturbative part
(Fig1 (b, c, d)),  the light quark condensates diagrams up to
operators having dimension $d=5$ i.e.,
 operators $d=3$, $<\overline{q}q>$, $d=4$, $m_{s}<\overline{q}q>$, $d=5$,
$m_{0}^{2}<\overline{q}q>$  are assumed. Contributions coming from
the light quark condensates diagrams are eliminated by applying the
double Borel transformations with respect to the initial and final
momentums $p$ and $p'$, so as first correction in the
nonperturbative part of the correlation function in the QCD side,
the two gluon condensates are calculated (see Fig. 2 (a, b, c, d, e,
f)).
 In calculating the bare-loop
contribution, we first write the double dispersion representation
for the coefficients of the corresponding Lorentz structures,
appearing in the correlation function, as:
\begin{equation}\label{10au}
\Pi_i^{per}=-\frac{1}{(2\pi)^2}\int ds'\int
ds\frac{\rho_{i}(s,s',q^2)}{(s-p^2)(s'-p'^2)}+\textrm{ subtraction
terms}.
\end{equation}
The integration region for the perturbative contribution
 in Eq. (\ref{10au}) is determined from the fact that arguments of the
 three $\delta$ functions must vanish simultaneously. The physical
 region in the  $s$ and $s'$ plane is described by the following
 inequalities:\\
 \begin{equation}\label{13au}
 -1\leq\frac{2ss'+(s+s'-q^2)(m_{b}^2-s-m_{c}^2)+(m_{c}^2-m_{q'}^2)2s}
 {\lambda^{1/2}(m_{b}^2,s,m_{c}^2)\lambda^{1/2}(s,s',q^2)}\leq+1.
\end{equation}
The spectral densities $\rho_{i}(s,s',q^2)$ can be calculated from
the usual Feynman integral with the help of Cutkosky rules, i.e., by
replacing the quark propagators with Dirac delta functions:
$\frac{1}{p^2-m^2}\rightarrow-2\pi i\delta(p^2-m^2),$ which implies
that all quarks are real. After standard calculations for the
corresponding spectral densities we obtain:
\begin{eqnarray}\label{11au}
\rho_{V}^{V-A}(s,s',q^2)&=&-N_{c}I_{0}(s,s',q^2)\left[-4{m_{c}+4(m_{b}-m_{c})B_{1}+4(m_{q'}-m_{c})B_{2}}\right],\nonumber\\
\rho_{0}^{V-A}(s,s',q^2)&=&-N_{c}I_{0}(s,s',q^2)[8(-m_{b}+m_{c})A_{1}-4m_{b}m_{c}m_{q'}\nonumber\\&+&
4(m_{q'}+ m_{b}-m_{c })m_{c}^2-2(m_{q'}-m_{c})\Delta\nonumber
\\&-&
2(m_{b}-m_{c})\Delta'-2m_{c}u],\nonumber \\
\rho_{+}^{V-A}(s,s',q^2)&=&N_{c}I_{0}(s,s',q^2)[4(m_{b}-m_{c})(A_{2}+A_{3})+2(m_{b}-3m_{c})B_{1}
\nonumber \\
 &&-2(m_{c}-m_{q'})B_{2}-2m_{c}]
,\nonumber \\
\rho_{-}^{V-A}(s,s',q^2)&=&N_{c}I_{0}(s,s',q^2)[4(m_{b}-m_{c})(A_{2}-A_{3})-2(m_{b}+m_{c})B_{1}
\nonumber \\
&& +2(m_{c}-m_{q'})B_{2} +2m_{c}],\nonumber \\
\rho_{V}^{T-PT}(s,s',q^2)&=&N_cI_0(s,s',q^2)[B_1(m_c^2-m_{b}m_{c}+m_b m_{q'}-m_c m_{q'}+s-\Delta)\nonumber\\
&-&B_2(m_c^2-m_bm_c+m_bm_{q'}-m_cm_{q'}+s'-\Delta')\nonumber
\\&-&(m_{q'}m_c+m_bm_c)],\nonumber \\
\rho_{0}^{T-PT}(s,s',q^2)&=&-N_cI_0(s,s',q^2)[2\Delta'(-m_c^2+m_bm_c-m_bm_{q'}+m_cm_{q'}+s)\nonumber\\
&-&2\Delta (-m_c^2+m_bm_c-m_bm_{q'}+m_cm_{q'}+s')\nonumber \\&+&4s
m_c(-m_{c}+m_{q'})+4s'm_c(m_c-m_b) +2u
m_c(-m_{q'}+m_b)\nonumber\\&+&8A_1(s-u/2)],\nonumber \\
\rho_{-}^{T-PT}(s,s',q^2)&=&-N_cI_0(s,s',q^2)[2B_1(2m_c^2-\Delta'+\Delta
-m_bm_c+m_bm_{q'}-m_cm_{q'}-s)\nonumber\\
&-&2B_2(2m_c^2-m_bm_c+m_bm_{q'}-m_cm_{q'}-s')\nonumber \\
&+&4(s-u/2)(A_2-A_3)+2m_c(m_b-m_{q'})],
\end{eqnarray}
 where
\begin{eqnarray}\label{12}
I_{0}(s,s',q^2)&=&\frac{1}{4\lambda^{1/2}(s,s',q^2)},\nonumber\\
 \lambda(a,b,c)&=&a^2+b^2+c^2-2ac-2bc-2ab,\nonumber \\
\Delta'&=&(s'+m_{c}^2 - m_{q'}^2),\nonumber\\
\Delta&= &(s+m_{c}^2 -
m_{b}^2),\nonumber\\
 u &=& s + s' - q^2,\nonumber\\
 B_{1}&=&\frac{1}{\lambda(s,s',q^2)}[2s'\Delta-\Delta'u],\nonumber\\
 B_{2}&=&\frac{1}{\lambda(s,s',q^2)}[2s\Delta'-\Delta u],\nonumber\\
 A_{1}&=&-\frac{1}{2\lambda(s,s',q^2)}[(4ss'm_c^2-s\Delta'^2
 -s'\Delta^2-u^2m_c^2+u\Delta\Delta')]
  ,\nonumber\\
 A_{2}&=&-\frac{1}{\lambda^2(s,s',q^2)}[8ss'^2m_c^2-2ss'\Delta'^2-6s'^2\Delta^2
 -2u^2s'm_c^2\nonumber\\
 &+&6s'u\Delta\Delta'-u^2\Delta'^2]
 ,\nonumber\\
 A_{3}&=&\frac{1}{\lambda^{2}(s,s',q^2)}[4ss'um_c^2+4ss'\Delta\Delta'
 -3su\Delta'^2
 -3u\Delta^2s'-u^3m_c^2+2u^2\Delta\Delta'].\nonumber\\
 \end{eqnarray}
The subscripts V, 0 and $\pm$ for $\rho^{V-A}$ correspond to the
coefficients of the structures proportional to
$i\varepsilon_{\mu\nu\alpha\beta}p^{\alpha}p'^{\beta}$,
  $g_{\mu\nu}$ and $\frac{1}{2}(p_{\mu}p_{\nu}
 \pm p'_{\mu}p_{\nu})$, respectively  and
  the subscripts V, 0 and $-$ for $\rho^{T-PT}$ correspond to the
coefficients of the structures proportional to
$\varepsilon_{\mu\nu\alpha\beta}p^{\alpha}p'^{\beta}$,
  $g_{\mu\nu}$ and $(p_{\mu}p_{\nu}
 - p'_{\mu}p_{\nu})$. In Eq. (\ref{11au}) $N_{c}=3$ is the number of
 colors.
Then, we consider the nonperturbative part of the correlator. As
we mentioned before, the light quark condensate
 contribution to the nonperturbative part of the correlation function is
 zero, so we calculate the gluon condensates diagrams shown in Fig. 2.
 The calculations of these diagrams
are performed in the Fock--Schwinger fixed--point gauge
\cite{R7320,R7321,R7322} \bea x^\mu A_\mu^a = 0~,\nnb \eea where
$A_\mu^a$ is the gluon field. In calculating the gluon condensate
contributions,   the following types of integrals  are appeared
\cite{Aliev3, R7323}: \bea \label{e7323} I_0[a,b,c] = \int
\frac{d^4k}{(2 \pi)^4} \frac{1}{\left[ k^2-m_b^2 \right]^a \left[
(p+k)^2-m_c^2 \right]^b \left[ (p^\prime+k)^2-m_{q'}^2\right]^c}~,
\nnb \\ \nnb \\
I_\mu[a,b,c] = \int \frac{d^4k}{(2 \pi)^4} \frac{k_\mu}{\left[
k^2-m_b^2 \right]^a \left[ (p+k)^2-m_c^2 \right]^b \left[
(p^\prime+k)^2-m_{q'}^2\right]^c}~,
\nnb \\ \nnb \\
I_{\mu\nu}[a,b,c] =\int \frac{d^4k}{(2 \pi)^4} \frac{k_\mu
k_\nu}{\left[ k^2-m_b^2 \right]^a \left[ (p+k)^2-m_c^2 \right]^b
\left[ (p^\prime+k)^2-m_{q'}^2\right]^c}~, \eea where $k$ is the
momentum of the spectator quark $c$. These integrals can be
calculated by continuing to Euclidean space--time and using
Schwinger representation for the Euclidean propagator \bea
\label{e7324} \frac{1}{k^2+m^2} = \frac{1}{\Gamma(\alpha)}
\int_0^\infty d\alpha \, \alpha^{n-1} e^{-\alpha(k^2+m^2)}~, \eea
which is very suitable for the Borel transformation since \bea
\label{e7325} {\cal B}_{\hat{p}^2} (M^2) e^{-\alpha p^2} = \delta
(1/M^2-\alpha)~. \eea Performing integration over loop momentum
and over the two parameters which we have used in the exponential
representation of propagators \cite{R7321}, and applying double
Borel transformations over $p^2$ and $p^{\prime 2}$, we get the
Borel transformed form of the integrals in Eq. (\ref{e7323}) (see
also \cite{R7321}):
\begin{figure}
\vspace*{-1cm}
\begin{center}
\includegraphics[width=11cm]{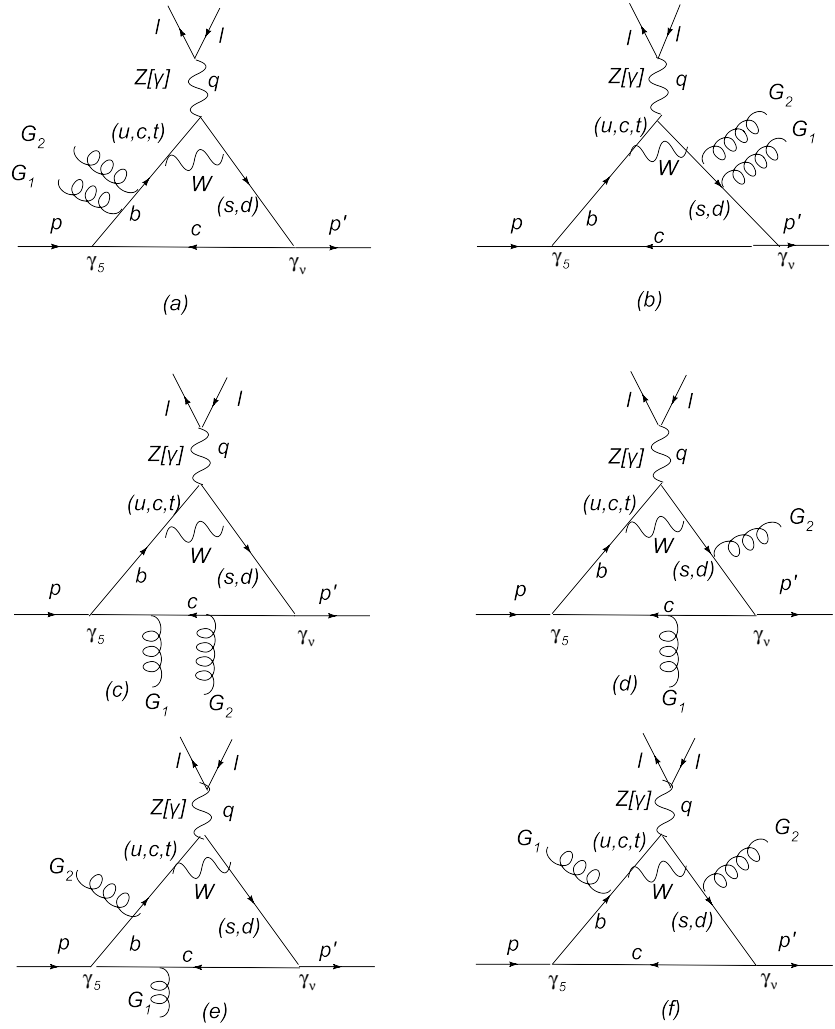}
\end{center}
\caption{Gluon condensate contributions to $B_c \rightarrow
D_{s,d}^{*}~l^+ l^-$ transitions  } \label{fig2}
\end{figure}
 \bea
\label{e7326} \hat{I}_0(a,b,c)& =&i\frac{(-1)^{a+b+c}}{16
\pi^2\,\Gamma(a) \Gamma(b) \Gamma(c)}
(M_1^2)^{2-a-b} (M_2^2)^{2-a-c} \, {\cal U}_0(a+b+c-4,1-c-b)~, \nnb \\ \nnb \\
\hat{I}_\mu(a,b,c) &=&\hat{I}_1(a,b,c) p_\mu +
\hat{I}_2(a,b,c) p'_\mu~, \nnb \\ \nnb \\
\hat{I}_{\mu\nu}(a,b,c)& =& \hat{I}_3(a,b,c) g_{\mu\nu} +
 \hat{I}_4(a,b,c)
p_\mu p_\nu
+ \hat{I}_5(a,b,c)  p'_\mu p'_\nu \nnb \\
&+&  \hat{I}_6 (a,b,c)  p_\mu p'_\nu + \hat{I}_7 (a,b,c)  p_\nu
p'_\mu, \eea where
\begin{eqnarray}
 \hat{I}_1(a,b,c) &=& i \frac{(-1)^{a+b+c+1}}{16
\pi^2\,\Gamma(a) \Gamma(b) \Gamma(c)}
(M_1^2)^{2-a-b} (M_2^2)^{3-a-c} \, {\cal U}_0(a+b+c-5,1-c-b)~, \nonumber \\ \nonumber \\
\hat{I}_2(a,b,c) &=& i \frac{(-1)^{a+b+c+1}}{16 \pi^2\,\Gamma(a)
\Gamma(b) \Gamma(c)}
(M_1^2)^{3-a-b} (M_2^2)^{2-a-c} \, {\cal U}_0(a+b+c-5,1-c-b)~, \nonumber \\ \nonumber \\
\hat{I}_3(a,b,c) &=& i \frac{(-1)^{a+b+c+1}}{32 \pi^2\,\Gamma(a)
\Gamma(b) \Gamma(c)}
(M_1^2)^{3-a-b} (M_2^2)^{3-a-c} \, {\cal U}_0(a+b+c-6,2-c-b)~,\nonumber \\ \nonumber \\
\hat{I}_4(a,b,c) &=& i \frac{(-1)^{a+b+c}}{16 \pi^2\,\Gamma(a)
\Gamma(b) \Gamma(c)}
(M_1^2)^{2-a-b} (M_2^2)^{4-a-c} \, {\cal U}_0(a+b+c-6,1-c-b)~,\nonumber \\ \nonumber \\
\hat{I}_5(a,b,c) &=& i \frac{(-1)^{a+b+c}}{16 \pi^2\,\Gamma(a)
\Gamma(b) \Gamma(c)}
(M_1^2)^{4-a-b} (M_2^2)^{2-a-c} \, {\cal U}_0(a+b+c-6,1-c-b)~,\nonumber \\ \nonumber \\
\hat{I}_6(a,b,c) &=& i \frac{(-1)^{a+b+c}}{16 \pi^2\,\Gamma(a)
\Gamma(b) \Gamma(c)} (M_1^2)^{3-a-b} (M_2^2)^{3-a-c} \, {\cal
U}_0(a+b+c-6,1-c-b)~,\nonumber \\ \nonumber \\
\hat{I}_7(a,b,c) &=&\hat{I}_6(a,b,c)~,
  \end{eqnarray}
  and $M_1^2$ and $M_2^2$ are the Borel
parameters. The function ${\cal U}_0(\alpha,\beta)$ is defined as
\bea {\cal U}_0(a,b) = \int_0^\infty dy (y+M_1^2+M_2^2)^a y^b
\,exp\left[ -\frac{B_{-1}}{y} - B_0 - B_1 y \right]~, \nnb \eea
where \bea \label{e7328} B_{-1}& =& \frac{1}{M_1^2M_2^2}
\left[m_{q'}^2M_1^4+m_b^2 M_2^4 + M_2^2M_1^2 (m_b^2+m_{q'}^2
-q^2) \right] ~, \nnb \\
B_0 &=& \frac{1}{M_1^2 M_2^2} \left[ (m_{q'}^2+m_c^2) M_1^2 + M_2^2
(m_b^2+m_c^2)
\right] ~, \nnb \\
B_{1} &=& \frac{m_c^2}{M_1^2 M_2^2}~. \eea Hat in above equations
denotes the  double Borel transformed form of integrals.

The QCD sum rules for the form factors $A'_{V}$, $A'_{0}$,
$A'_{+}$ , $A'_{-}$, $T'_{V}$, $T'_{0}$ and $T'_{-}$ are obtained
by equating the phenomenological expression given in
Eq.~(\ref{9amplitude}) and QCD side in  Eq.~(\ref{QCD side}) and
applying double Borel transformations with respect to the
variables $p^2$ and $p'^2$ ($p^2\rightarrow
M_{1}^2,p'^2\rightarrow M_{2}^2$) in order to suppress the
contributions of higher states and continuum:
\begin{eqnarray}\label{15au}
A'_{i}(q^2)&=&\frac{(m_{b}+m_{c})
}{f_{B_{c}}m_{B_{c}}^2}\frac{1}{f_{D_{q'}^{*}}m_{D_{q'}^{*}}}~e^{m_{B_{c}}^2/M_{1}^2}
e^{m_{D_{q'}^{*}}^2/M_{2}^2}
\left[\vphantom{\int_0^{x_2}}\frac{1}{(2\pi)^2}\int_{(m_c+m_{q'})^2}^{s'_0}
ds'
 \int_{f_{-}(s')}^{min(s_0,f_{+}(s'))} ds\rho_{i}^{V-A}(s,s',q^2)e^{-s/M_{1}^2-s'/M_{2}^2}\right.\nonumber
\\&+&\left.i\frac{1}{24\pi^2}{C^{A'_{i}}}<\frac{\alpha_{s}}{\pi}G^{2}>\vphantom{\int_0^{x_2}}\right],\nonumber\\
T'_{i}(q^2)&=&\frac{(m_{b}+m_{c})
}{f_{B_{c}}m_{B_{c}}^2}\frac{1}{f_{D_{q'}^{*}}m_{D_{q'}^{*}}}~e^{m_{B_{c}}^2/M_{1}^2}
e^{m_{D_{q'}^{*}}^2/M_{2}^2}
\left[\vphantom{\int_0^{x_2}}\frac{1}{(2\pi)^2}\int_{(m_c+m_{q'})^2}^{s'_0}
ds'
 \int_{f_{-}(s')}^{min(s_0,f_{+}(s'))} ds\rho_{i}^{T-PT}(s,s',q^2)e^{-s/M_{1}^2-s'/M_{2}^2}\right.\nonumber
\\&+&\left.i\frac{1}{24\pi^2}{C^{T'_{i}}}<\frac{\alpha_{s}}{\pi}G^{2}>\vphantom{\int_0^{x_2}}\right],\nonumber\\
 \end{eqnarray}
where  coefficients $C^{A'_{i}}$ and $C^{T'_{i}}$ come from the
gluon condensates contributions and their explicit expressions are
given in the appendix. The $s_{0}$ and $s'_{0}$ are the continuum
thresholds in $s$ and $s'$ channels, respectively and $f_{\pm}(s')$
in the lower and upper limit in the integration over $s$ are
calculated from inequality ( \ref{13au}) with respect to $s$, i.e.,
$s=f_{\pm}(s')$. By $min(s_0,f_{+}(s'))$, for each value of the
$q^{2}$, we select the smaller one between $s_{0}$ and $f_{+}$. In
Eq.~(\ref{15au}), in order to subtract the contributions of the
higher states and the continuum the quark-hadron duality assumption
is also
 used, i.e., it is assumed
that
\begin{eqnarray}
\rho^{higher states}(s,s') = \rho^{OPE}(s,s') \theta(s-s_0)
\theta(s-s'_0).
\end{eqnarray}
In three point QCD sum rules in the case of double dispersion
integrals,  for $q^2>0$ values, their could be a deviation between
double dispersion integrals in Eq. (\ref{15au}) and corresponding
coefficients  of the structures in the feynman amplitudes in the
bare loop diagram (see Fig. 1a), i.e., for our case
\begin{eqnarray}
\int\frac{d^4k}{(2\pi)^4}\frac{Tr\{(\not\!k+m_{c})\gamma_{\nu}(\not\!p'+\not\!k+m_{q'})\gamma_{\mu}(1-\gamma_{5})(\not\!p+\not\!k+m_{b})\}}{[k^{2}-m_{c}^{2}]
[(p'+k)^{2}-m_{q'}^{2}][(p+k)^{2}-m_{b}^{2}]},
\end{eqnarray}
 where k is the momentum of the spectator quark c. In those cases, the double spectral density receives contributions beyond the contributions due to Landau-type singularities.
 This problem has been studied  in details in \cite{problem}. In our
 case, we have
studied such contributions and with the above continuum subtraction
and the selecting integration region,   they turn out to be small.
Here, we neglect such contributions, so our calculations are
trustworthy in the above integration region.

 In our calculations the following Borel transformations are also
used
\begin{eqnarray}
{\cal B}_{p^{2}}\left\{\frac{1}{m^2(s)-p^2}\right\}& =&
e^\frac{m^2(s)}{M_{1}^2},\nonumber
\\
{\cal B}_{p'^2}\left\{\frac{1}{m^2(s')-p'^2}\right\}& =&
e^\frac{m^2(s')}{M_{2}^2}.
\end{eqnarray}
\section{Numerical analysis}
From the explicit expressions for the form factors  $A_{V}$,
$A_{0}$, $A_{+}$ , $A_{-}$, $T_{1}$, $T_{2}$, $T_{3}$ and effective
hamiltonian, it is clear that the main input parameters entering to
the expressions are gluon condensate, Wilson coefficients
$C_7^{eff}$, $C_9^{eff}$ and $C_{10}$ , the CKM matrix elements
$V_{tb}$, $V_{ts}$ and $V_{td}$, leptonic decay constants
$f_{B_{C}}$ and $f_{D_{s}^{\ast}}$, Borel parameters $M_{1}^2$ and
$M_{2}^2$, as well as the continuum thresholds $s_{0}$ and $s'_{0}$.

In further numerical analysis we choose the value of the gluon
condensate  $<\frac{\alpha_{s}}{\pi}G^{2}>=0.012~ GeV ^{4}$
\cite{Shifman1},
 $C_7^{eff}=-0.313$, $C_9^{eff}=4.344$, $C_{10}=-4.669$
 \cite{Buras,Bashiry},  $\mid
V_{tb}\mid=0.77^{+0.18}_{-0.24}$, $\mid
V_{ts}\mid=(40.6\pm2.7)\times10^{-3}$ $\mid
V_{td}\mid=(7.4\pm0.8)\times10^{-3}$ \cite {Ceccucci},
$f_{D_{s}^{\ast}} =266\pm32
  ~MeV $ \cite{Colangelo1}, $f_{D^{\ast}}=0.23\pm0.02~GeV$
\cite{Bowler}, $f_{B_{c}} =350\pm10
  ~MeV $ \cite{Colangelo2, Kiselev, Alievy}, $ m_{c}(\mu=m_{c})=
 1.275\pm
 0.015~ GeV$, $m_{s}(1~ GeV)\simeq 142 ~MeV$ \cite{Huang} ,  $m_{b} =
(4.7\pm
 0.1)~GeV$ \cite{Ioffe} , $m_{d}=(3-7)~MeV$, $m_{D_{s}^{\ast}}=2.112~GeV$, $m_{D^{\ast}}=2.010~GeV$
  and $ m_{B_{C}}=6.258~GeV$ \cite{Yao}.

The expressions for the form factors  contain four auxiliary
parameters: Borel mass squares $M_{1}^2$ and $M_{2}^2$ and continuum
threshold $s_{0}$  and $s'_{0}$. These are not physical quantities,
hence the physical quantities , form factors, must be independent of
these auxiliary parameters. In other words, we should find the
"working regions" of these auxiliary parameters, where the form
factors are independent of them. We try to find the working region
of $M_{1}^2$ and $M_{2}^2$ by requiring that the upper bound of
$M_{1,2}^{2}$ is fixed such that the continuum contribution should
be less than the contribution of the first resonance. The lower
bound of $M_{1,2}^{2}$ are determined by requiring that the highest
power of $1/M_{1,2}^{2}$ is less than about $30^{0}/_{0}$ of  the
highest power of $M_{1,2}^{2}$. These two conditions are both
satisfied in the  following regions; $10~GeV^2\leq
M_{1}^{2}\leq25~GeV^2 $ and $4~GeV^2\leq M_{2}^{2}\leq10~GeV^2 $.
The continuum threshold parameters are also determined from the
two-point QCD sum rules: $s_{0} = 45 ~GeV ^{2}$ and $s'_{0} =8~
GeV^2$ \cite{Aliev1, Colangelo1, Shifman1}.

 In order to estimate the decay width of $B_c \rightarrow
D_{s,d}^{*}~l^+ l^-$ decays, it is necessary to know the $q^2$
dependency of the form factors  $A_{V}$, $A_{0}$, $A_{+}$ , $A_{-}$,
$T_{1}$, $T_{2}$ and  $T_{3}$ in the whole physical region, $
4m_{l}^2 \leq q^2 \leq (m_{B_{c}} - m_{D_{s,d}^{*}})^2$, which is
correspond to $ 4m_{l}^2 \leq q^2 \leq 17.2~GeV^2$ and $ 4m_{l}^2
\leq q^2 \leq 18~GeV^2$ for $D_{s}^{*}$ and $D_{d}^{*}$,
respectively. For extracting the $q^2$ dependencies of the form
factors from QCD sum rules, we should consider a range of $ q^2$
where the correlation function can reliably be calculated. Our sum
rules for the form factors are truncated at $q^2 \simeq14 ~GeV^2$.
In order to extend our results to the full physical region, we look
for parametrization of the form factors in such a way that in the
region $0 \leq q^2 \leq 14~ GeV^2$, this parametrization coincides
with the sum rules prediction. The values of  the form factors at
$q^2=0$ are shown in Table I.

\begin{table}[h]
\centering
\begin{tabular}{|c|c|c|
} \hline
  &$ B_c \rightarrow D^*_{s} ~l^+ l^-$  & $B_c \rightarrow D^*_d ~l^+ l^-$\\ \cline{1-3}
 $A_{V}(0)$ &$0.54\pm0.018$&$0.63\pm0.055 $  \\\cline{1-3}
 $A_{0}(0)$ &$ 0.30\pm 0.017$&$0.34\pm0.005 $  \\\cline{1-3}
 $A_{+}(0)$ &$0.36\pm0.013$&$ 0.41\pm0.037$   \\\cline{1-3}
$A_{-}(0)$ & $-0.57\pm0.04$&$ -0.68\pm0.082 $  \\\cline{1-3}
 $T_{1}(0)$ &$0.31\pm0.017$&$ 0.36\pm0.039 $  \\\cline{1-3}
$T_{2}(0)$ & $0.33\pm0.016$&$ 0.37\pm0.039 $ \\\cline{1-3}
 $T_{3}(0)$ &$0.29\pm0.034$&$0.37\pm0.058$  \\\cline{1-3}
 \end{tabular}
 \vspace{0.8cm}
\caption{The values of  the form factors at $q^2=0$, for
$M_{1}^2=17~GeV^2$ and  $M_{2}^2=7~GeV^2$ .}\label{tab:1}
\end{table}

 \begin{table}[h]
\centering
\begin{tabular}{|c|c|c|c|c|c|c|c|} \hline
 &$A_{V}(q^2)$  & $A_{0}(q^2)$ &$A_{+}(q^2)$&$A_{-}(q^2)$&$T_{1}(q^2)$&$T_{2}(q^2)$&$T_{3}(q^2)$\\\cline{1-8}
 F(0) & 0.54 & 0.30&0.36&-0.57&0.31&0.33&0.29\\\cline{1-8}
 $ \alpha$ &-1.28  & -0.13&-0.67&-1.11&-1.28&-0.10&-0.91\\\cline{1-8}
$ \beta$ &-0.23 &-0.18&-0.066&-0.14&-0.23&-0.097&0.007\\\cline{1-8}
 \end{tabular}
 \vspace{0.8cm}
\caption{Parameters appearing in the form factors of the
$B_{c}\rightarrow D_{s}^{*}l^+ l^-$}decay in a four-parameter fit,
for $M_{1}^2=17~GeV^2$ and  $M_{2}^2=7~GeV^2$ \label{tab:1}
\end{table}
\begin{table}[h]
\centering
\begin{tabular}{|c|c|c|c|c|c|c|c|} \hline
&$A_{V}(q^2)$  & $A_{0}(q^2)$
&$A_{+}(q^2)$&$A_{-}(q^2)$&$T_{1}(q^2)$&$T_{2}(q^2)$&$T_{3}(q^2)$\\\cline{1-8}
 F(0) & 0.63& 0.34&0.41&-0.68&0.36&0.37&0.37\\\cline{1-8}
 $ \alpha$ &-1.22  & 0.015&-0.58&-1.06&-1.22&0.025&-0.90\\\cline{1-8}
$ \beta$ &-0.23 &-0.074&-0.022&-0.13&-0.23&0.002&0.014\\\cline{1-8}
 \end{tabular}
 \vspace{0.8cm}
\caption{Parameters appearing in the form factors of the
$B_{c}\rightarrow D_{d}^{*}l^+ l^-$}decay in a four-parameter fit,
for $M_{1}^2=17~GeV^2$ and $M_{2}^2=7~GeV^2$ \label{tab:2}
\end{table}

Our numerical calculations shows that the best parameterization of
the
form factors with respect to $q^2$ are as follows:\\
 \begin{equation}\label{17au}
 F(q^2)=\frac{F(0)}{1+ \alpha\hat{q}+\beta\hat{q}^{2}}
\end{equation}
where $\hat{q}=q^2/m_{B_{c}}^2$. The values of the parameters
 $F(0)$, $\alpha$ and $\beta$ for $B_{c}\rightarrow D_{s}^{*}l^+ l^-$ and $B_{c}\rightarrow D_{d}^{*}l^+ l^-$ are
given in  Tables II and III, respectively.

The next step is to calculate the total decay width. After straight
forward calculations, the differential decay width for these decays
are obtained
\begin{equation}\label{e6308}
\frac{d\Gamma}{d\hat{q}} = \frac{G_{F}^2 \alpha^2 m_{B_c}}{2^{14}
\pi^5} | V_{tb}V_{tq'}^{*} |^2
\lambda^{1/2}(1,\hat{r},\hat{q}) v \Delta(\hat{q})~,\nonumber \\
\end{equation}
where
\begin{eqnarray}
\Delta &=& \frac{2}{3 \hat{r} \hat{q}} m_{B_c}^2
 Re\Big[ - 12 m_{B_c}^2 \hat{m}_l^2 \hat{q} \lambda(1,\hat{r},\hat{q}) \Big\{
(E_3-D_2-D_3) E_1^\ast\nonumber \\ & -& (E_2+E_3-D_3) D_1^\ast
\Big\} +12 m_{B_c}^4 \hat{m}_l^2 \hat{q} (1-\hat{r})
\lambda(1,\hat{r},\hat{q})
(E_2-D_2) (E_3^\ast-D_3^\ast) \nonumber \\
&+& 48 \hat{m}_l^2 \hat{r} \hat{q} \Big\{ 3 E_1 D_1^\ast + 2
m_{B_c}^4 \lambda(1,\hat{r},\hat{q}) E_0 D_0^\ast \Big\} - 16
m_{B_c}^4 \hat{r}\hat{q} (\hat{m}_l^2-\hat{q})
\lambda(1,\hat{r},\hat{q})
\Big\{ | E_0|^2 + | D_0|^2 \Big\} \nonumber \\
&-& 6 m_{B_c}^4 \hat{m}_l^2 \hat{q} \lambda(1,\hat{r},\hat{q})
\Big\{ 2 (2+2\hat{r}-\hat{q}) E_2 D_2^\ast -
\hat{q} | (E_3-D_3)|^2 \Big\} \nonumber \\
&- &4 m_{B_c}^2 \lambda(1,\hat{r},\hat{q}) \Big\{ \hat{m}_l^2 (2 - 2
\hat{r} + \hat{q} ) + \hat{q} (1 - \hat{r} - \hat{q} ) \Big\}
(E_1 E_2^\ast + D_1 D_2^\ast) \nonumber\\
&+& \hat{q} \Big\{ 6 \hat{r} \hat{q} (3+v^2) +
\lambda(1,\hat{r},\hat{q}) (3-v^2) \Big\} \Big\{ | E_1|^2 + | D_1|^2
\Big\}\nonumber\\&-& 2 m_{B_c}^4 \lambda(1,\hat{r},\hat{q}) \Big\{
\hat{m}_l^2 [\lambda(1,\hat{r},\hat{q}) - 3 (1-\hat{r})^2] -
\lambda(1,\hat{r},\hat{q}) \hat{q} \Big\}
\Big\{ | E_2|^2 + | D_2|^2 \Big\}\Big]~, \nonumber\\
\end{eqnarray}

 and $\hat{q}=q^2/m_{B_c}^2$,
$\hat{r}=m_{D_{q'}^{*}}^2/m_{B_c}^2$, $\hat{m}_l=m_l/m_{B_c}$ and
$v=\sqrt{1-4\hat{m}_l^2/\hat{q}}$ is the final lepton velocity. We
have also used the  following definitions:
\begin{eqnarray} \label{e6307}
D_0 &=& (C_9^{eff}+C_{10})
\frac{A_{V}(q^2)}{m_{B_c}+m_{D_{q'}^\ast}} +
 (2m_bC_7^{eff}) \frac{T_{1}(q^2)}{q^2} ~, \nonumber\\
D_1 &=& (C_9^{eff}+C_{10}) (m_{B_c}+m_{D_{q'}^\ast}) A_{0}(q^2) +
(2m_bC_7^{eff}) (m_{B_c}^2-m_{D_{q'}^\ast}^2)
\frac{T_{2}(q^2)}{q^2} ~, \nonumber\\
D_2 &=& \frac{C_9^{eff}+C_{10}}{m_{B_c}+m_{D_{q'}^\ast}} A_{+}(q^2)
+(2m_bC_7^{eff}) \frac{1}{q^2}  \left[ T_{2}(q^2) +
\frac{q^2}{m_{B_c}^2-m_{D_{q'}^\ast}^2} T_{3}(q^2) \right]~,
\nonumber\\
D_3 &=& (C_9^{eff}+C_{10})\frac{A_{-}(q^2)}{m_{B_c}+m_{D_{q'}^\ast}}
-
 (2m_bC_7^{eff}) \frac{T_{3}(q^2)}{q^2} ~, \nonumber\\
E_0 &=& (C_9^{eff}-C_{10})
\frac{A_{V}(q^2)}{m_{B_c}+m_{D_{q'}^\ast}} +
 (2m_{b}C_7^{eff}) \frac{T_{1}(q^2)}{q^2} ~, \nonumber\\
E_1 &=& (C_9^{eff}-C_{10}) (m_{B_c}+m_{D_{q'}^\ast}) A_{0}(q^2) +
(2m_{b}C_7^{eff}) (m_{B_c}^2-m_{D_{q'}^\ast}^2)
\frac{T_{2}(q^2)}{q^2} ~, \nonumber\\
E_2 &=& \frac{C_9^{eff}-C_{10}}{m_{B_c}+m_{D_{q'}^\ast}} A_{+}(q^2)
+(2m_{b}C_7^{eff}) \frac{1}{q^2}  \left[ T_{2}(q^2) +
\frac{q^2}{m_{B_c}^2-m_{D_{q'}^\ast}^2} T_3(q^2) \right]~,
\nonumber\\
E_3 &=& (C_9^{eff}-C_{10})\frac{A_{-}(q^2)}{m_{B_c}+m_{D_{q'}^\ast}}
-
 (2m_{b}C_7^{eff}) \frac{T_3(q^2)}{q^2} . \nonumber\\
 \end{eqnarray}
At the end of this section, we present the value of the branching
ratio for $B_c \rightarrow D_{s,d}^{*}~l^+ l^-$ decays. Taking into
account the $q^{2}$ dependence of the form factors and performing
integration over $q^{2}$ in the limit $ 4m_{l}^{2}<q^{2}<
(m_{B_c}-m_{D_{s,d}^{*}} )^{2}$ and using the total life-time
$\tau_{B_{c}}=0.46 \times10^{-12}s$ \cite{Beneke}, we obtain  the
branching ratios for $B_{c}\rightarrow D_{s}^{*}l^+ l^-$ and
$B_{c}\rightarrow D_{d}^{*}l^+ l^-$ presented in Tables IV and V,
respectively. These  Tables  contain also  comparison of our results
with the predictions of the quark models. These  Tables show a good
agreement between our results and that of the quarks models
especially when we consider the errors.

In summary, we analyzed the $B_{c}\rightarrow D_{q'}^{\ast}ll$
transitions in the framework of the QCD sum rules. The $q^2$
dependent expressions for form factors were calculated. The quark
condensates contributions to the correlation function were zero, so
we considered the gluon corrections to the correlator. Finally, we
calculated the total decay width and branching ratio of these decays
and compared our results with the predictions of the quark models.
Our results are in good agreement with the quarks model.
\begin{table}[h]
\centering
\begin{tabular}{|c|c|c|c|} \hline
  & $B_{c}\rightarrow D_{s}^{\ast}\mu^{+}\mu^{-}$ &  $B_{c}\rightarrow D_{s}^{\ast}\tau^{+}\tau^{-}$
   & $B_{c}\rightarrow D_{s}^{\ast}e^{+}e^{-}$\\\cline{1-4}
 Present study & $(2.99\pm0.50)\times10^{-7}$& $(0.205\pm0.076)\times10^{-7}$& $(4.21\pm0.62)\times10^{-7}$\\\cline{1-4}
 \cite{tt1} & $(1.41-1.76)\times10^{-7}$  & $(0.15-0.22)\times10^{-7}$ & $-$\\\cline{1-4}
   \cite{tt2} &$ (3.14-4.09)\times10^{-7} $&$(0.34-0.51)\times10^{-7}$& $-$\\\cline{1-4}
  \end{tabular}
 \vspace{0.8cm}
\caption{Values for the branching ratio of  the  $B_{c}\rightarrow
D_{s}^{*}l^+ l^-$ decay and their comparison with the predictions of
the RCQM \cite{tt1} and LFQM (CQM) \cite{tt2}.} \label{tab:1}
\end{table}
\begin{table}[h]
\centering
\begin{tabular}{|c|c|c|c|} \hline
  &$B_{c}\rightarrow D^{\ast}_{d}\mu^{+}\mu^{-}$&$B_{c}\rightarrow D^{\ast}_{d}\tau^{+}\tau^{-}$
   &$B_{c}\rightarrow D^{\ast}_{d}e^{+}e^{-}$\\\cline{1-4}
 Present study &  $(1.58\pm0.20)\times10^{-8}$
 & $(0.156\pm0.013)\times10^{-8}$ &$(1.97\pm0.20)\times10^{-8}$\\\cline{1-4}
 \cite{tt1} &$(0.58-0.71)\times10^{-8} $& $(0.08-0.11)\times10^{-8}$& $-$\\\cline{1-4}
   \cite{tt2} &$(0.78-1.01)\times10^{-8} $& $(0.13-0.18)\times10^{-8}$& $-$\\\cline{1-4}
  \end{tabular}
 \vspace{0.8cm}
\caption{Values for the branching ratio of the $B_{c}\rightarrow
D_{d}^{*}l^+ l^-$ decay and their comparison with the predictions of
the RCQM \cite{tt1} and  LFQM (CQM) \cite{tt2}.} \label{tab:1}
\end{table}
\section{Acknowledgment}
  The authors would like to thank T. M. Aliev and A. Ozpineci  for their useful discussions. One of the authors (K. A) thanks  TUBITAK,
  Turkish scientific and research council, for their
  financial
  support.

\section*{Appendix}
In this section, we present the explicit expressions of the
coefficients $C^{A'_{i}}$ and $C^{T'_{i}}$ corresponding to the
gluon condensates entering to the expressions for the form factors
in Eq. (\ref{15au}). Here, we have ignored the s and d quark
masses.
\begin{eqnarray*}
C^{A'_{V}} &=&-30\,\mathit{I_{1}}(4,1,1){m_{{c}}}^{3}+10\,\mathit{I_{2}}(3,2,1)%
{m_{{c}}}^{3}-20\,\mathit{I_{1}}^{{0,1}}(3,1,2)m_{{b}}+20\,\mathit{I_{1}}^{{%
0,1}}(3,1,2)m_{{c}} \\
&&+10\,\mathit{I_{1}}^{{0,2}}(3,2,2)m_{{b}}-20\,\mathit{I_{1}}(2,3,1){m_{{b}}%
}^{3}+20\,\mathit{I_{2}}^{{0,1}}(3,2,2){m_{{c}}}^{3}+20\,\mathit{I_{1}}^{{0,1%
}}(3,2,2){m_{{c}}}^{3} \\
&&-20\,\mathit{I_{1}}^{{0,1}}(3,2,1)m_{{b}}+20\,\mathit{I_{1}}^{{0,1}%
}(2,2,2)m_{{c}}-20\,\mathit{I_{1}}^{{0,1}}(2,2,2)m_{{b}}+20\,\mathit{I_{1}}%
(1,2,2)m_{{b}} \\
&&-20\,\mathit{I_{1}}(1,2,2)m_{{c}}-10\,\mathit{I_{1}}^{{0,1}}(3,2,2){m_{{b}}%
}^{3}-20\,\mathit{I_{1}}(2,2,2){m_{{c}}}^{3}-10\,\mathit{I_{0}}(3,2,2){m_{{c}%
}}^{5} \\
&&+10\,\mathit{I_{1}}(2,2,2){m_{{b}}}^{3}+20\,\mathit{I_{0}}(2,2,1)m_{{c}%
}-10\,\mathit{I_{1}}(3,2,2){m_{{c}}}^{5}-10\,{m_{{c}}}^{3}\mathit{I_{0}}%
(3,1,2) \\
&&+30\,m_{{c}}\mathit{I_{0}}^{{0,1}}(3,1,2)-10\,m_{{c}}\mathit{I_{0}}^{{0,2}%
}(3,2,2)+60\,m_{{b}}\mathit{I_{2}}(1,3,1)+10\,m_{{c}}\mathit{I_{2}}^{{0,1}%
}(3,2,1) \\
&&-20\,m_{{c}}\mathit{I_{0}}(1,2,2)-20\,{m_{{c}}}^{3}\mathit{I_{0}}%
(2,2,2)+20\,m_{{c}}\mathit{I_{0}}^{{0,1}}(2,2,2)-60\,\mathit{I_{1}}(1,4,1){%
m_{{b}}}^{3} \\
&&-20\,\mathit{I_{1}}(2,1,2)m_{{c}}+30\,\mathit{I_{1}}(2,1,2)m_{{b}}-30\,{m_{%
{c}}}^{3}\mathit{I_{2}}(4,1,1)+20\,m_{{c}}\mathit{I_{2}}^{{0,1}}(2,2,2) \\
&&-20\,{m_{{c}}}^{3}\mathit{I_{2}}(2,2,2)+10\,m_{{c}}\mathit{I_{2}}%
(3,1,1)+100\,m_{{b}}\mathit{I_{0}}(1,3,1)+60\,{m_{{b}}}^{2}m_{{c}}\mathit{%
I_{0}}(1,4,1) \\
&&+40\,m_{{b}}{m_{{c}}}^{2}\mathit{I_{0}}(2,3,1)+60\,{m_{{b}}}^{2}m_{{c}}%
\mathit{I_{2}}(1,4,1)-10\,m_{{c}}\mathit{I_{1}}(3,1,1)-10\,m_{{c}}\mathit{%
I_{2}}^{{0,2}}(3,2,2) \\
&&-30\,{m_{{c}}}^{3}\mathit{I_{0}}(4,1,1)+20\,m_{{b}}\mathit{I_{1}}%
(1,3,1)+40\,m_{{b}}\mathit{I_{2}}^{{0,1}}(2,3,1)-10\,{m_{{c}}}^{3}\mathit{%
I_{2}}(3,1,2) \\
&&-10\,m_{{c}}\mathit{I_{0}}(3,1,1)+20\,m_{{b}}\mathit{I_{1}}^{{0,1}%
}(2,3,1)-30\,m_{{c}}\mathit{I_{0}}(2,1,2)-30\,m_{{c}}\mathit{I_{2}}(2,1,2) \\
&&+20\,m_{{c}}\mathit{I_{0}}^{{0,1}}(3,2,1)+30\,m_{{c}}\mathit{I_{2}}^{{0,1}%
}(3,1,2)-20\,m_{{c}}\mathit{I_{2}}(1,2,2)+10\,\mathit{I_{0}}^{{0,1}}(3,2,2){%
m_{{b}}}^{2}m_{{c}} \\
&&+10\,\mathit{I_{2}}(3,2,2){m_{{c}}}^{3}{m_{{b}}}^{2}-30\,\mathit{I_{1}}%
(3,2,1){m_{{b}}}^{2}m_{{c}}+10\,\mathit{I_{1}}(3,2,1)m_{{b}}{m_{{c}}}%
^{2}+30\,\mathit{I_{1}}(4,1,1)m_{{b}}{m_{{c}}}^{2} \\
&&-20\,\mathit{I_{2}}(3,2,1){m_{{b}}}^{2}m_{{c}}+10\,\mathit{I_{1}}(3,2,2){%
m_{{c}}}^{3}{m_{{b}}}^{2}+10\,\mathit{I_{1}}(3,2,2)m_{{b}}{m_{{c}}}^{4}-10\,%
\mathit{I_{1}}(3,2,2){m_{{b}}}^{3}{m_{{c}}^{2}} \\
&&+60\,\mathit{I_{1}}(1,4,1){m_{{b}}}^{2}m_{{c}}+10\,\mathit{I_{1}}^{{0,1}%
}(3,2,2){m_{{b}}}^{2}m_{{c}}-20\,\mathit{I_{1}}^{{0,1}}(3,2,2)m_{{b}}{m_{{c}}%
}^{2}-10\,\mathit{I_{2}}(3,2,1)m_{{b}}{m_{{c}}^{2}} \\
&&+10\,\mathit{I_{2}}^{{0,1}}(3,2,2){m_{{b}}}^{2}m_{{c}}+20\,\mathit{I_{1}}%
(2,3,1)m_{{b}}{m_{{c}}}^{2}-20\,\mathit{I_{0}}(3,2,1){m_{{b}}}^{2}m_{{c}%
}+10\,\mathit{I_{0}}(3,2,1)m_{{b}}{m_{{c}}}^{2} \\
&&+10\,\mathit{I_{0}}(3,2,2){m_{{c}}}^{3}{m_{{b}}}^{2}+20\,\mathit{I_{1}}%
(2,2,2)m_{{b}}{m_{{c}}}^{2}+20\,\mathit{I_{1}}(2,2,1)m_{{c}}-50\,\mathit{%
I_{1}}(2,2,1)m_{{b}} \\
&&-20\,\mathit{I_{0}}(2,2,1)m_{{b}}+20\,\mathit{I_{0}}^{{0,1}}(3,2,2){m_{{c}}%
}^{3}+30\,\mathit{I_{2}}(2,2,1)m_{{c}}-10\,\mathit{I_{2}}(3,2,2){m_{{c}}}^{5}
\\
&&-20\,\mathit{I_{2}}(2,2,1)m_{{b}}+20\,\mathit{I_{1}}(3,2,1){m_{{b}}}%
^{3}-10\,\mathit{I_{1}}^{{0,2}}(3,2,2)m_{{c}}+20\,\mathit{I_{1}}^{{0,1}%
}(3,2,1)m_{{c}}
\end{eqnarray*}
\begin{eqnarray*}
C^{A'_{0}} &=&-5\,\mathit{I_{0}}(2,2,1){m_{{b}}}^{3}-5\,\mathit{I_{0}}(3,1,1){%
m_{{c}}}^{3}-40\,\mathit{I_{3}}^{{0,1}}(3,2,2){m_{{c}}}^{3}+20\,\mathit{I_{3}%
}^{{0,1}}(3,2,2){m_{{b}}}^{3} \\
&&-10\,\mathit{I_{0}}(1,3,1)m_{{b}}{m_{{c}}}^{2}-15\,\mathit{I_{0}}^{{0,1}%
}(3,2,2)m_{{b}}{m_{{c}}}^{4}+5\,\mathit{I_{0}}^{{0,1}}(3,2,2)m_{{c}}{m_{{b}}}%
^{4}+15\,\mathit{I_{0}}^{{0,2}}(3,2,2)m_{{b}}{m_{{c}}}^{2} \\
&&-5\,\mathit{I_{0}}^{{0,2}}(3,2,2)m_{{c}}{m_{{b}}}^{2}-20\,\mathit{I_{3}}%
(3,2,2){m_{{c}}}^{3}{m_{{b}}}^{2}-10\,\mathit{I_{0}}(2,3,1){m_{{b}}}^{3}{m_{{%
c}}}^{2}+10\,\mathit{I_{0}}(2,3,1)m_{{b}}{m_{{c}}}^{4} \\
&&-20\,\mathit{I_{3}}(3,1,2)m_{{b}}{m_{{c}}}^{2}-20\,\mathit{I_{3}}(2,2,2)m_{%
{b}}{m_{{c}}}^{2}+20\,\mathit{I_{3}}(2,2,2)m_{{c}}{m_{{b}}}^{2}+40\,\mathit{%
I_{3}}(2,3,1)m_{{b}}{m_{{c}}}^{2} \\
&&+30\,\mathit{I_{0}}(1,3,1)m_{{c}}{m_{{b}}}^{2}-10\,\mathit{I_{0}}(2,2,2){%
m_{{c}}}^{3}{m_{{b}}}^{2}+10\,\mathit{I_{0}}(2,2,1)m_{{c}}{m_{{b}}}^{2}-15\,%
\mathit{I_{0}}(2,2,1)m_{{b}}{m_{{c}}}^{2} \\
&&+10\,\mathit{I_{0}}(3,2,1){m_{{b}}}^{3}{m_{{c}}}^{2}-20\,\mathit{I_{3}}%
(3,2,2)m_{{b}}{m_{{c}}}^{4}+20\,\mathit{I_{3}}(3,2,2){m_{{b}}}^{3}{m_{{c}}}%
^{2}-20\,\mathit{I_{0}}^{{0,1}}(2,3,1)m_{{b}}{m_{{c}}}^{2} \\
&&+10\,\mathit{I_{0}}^{{0,1}}(2,2,2)m_{{c}}{m_{{b}}}^{2}-30\,\mathit{I_{0}}^{%
{0,1}}(2,2,2)m_{{b}}{m_{{c}}}^{2}+15\,\mathit{I_{0}}(4,1,1)m_{{b}}{m_{{c}}}%
^{4}-15\,\mathit{I_{0}}(4,1,1){m_{{c}}}^{3}{m_{{b}}}^{2} \\
&&-15\,\mathit{I_{0}}(2,1,2)m_{{c}}{m_{{b}}}^{2}+15\,\mathit{I_{0}}(2,1,2)m_{%
{b}}{m_{{c}}}^{2}+10\,\mathit{I_{0}}(3,1,1)m_{{b}}{m_{{c}}}^{2}+10\,\mathit{%
I_{0}}^{{0,1}}(3,2,2){m_{{c}}}^{3}{m_{{b}}}^{2} \\
&&-10\,\mathit{I_{0}}(1,2,2)m_{{c}}{m_{{b}}}^{2}+20\,\mathit{I_{0}}(1,2,2)m_{%
{b}}{m_{{c}}}^{2}-120\,\mathit{I_{3}}(1,4,1)m_{{c}}{m_{{b}}}^{2}-40\,\mathit{%
I_{3}}(3,2,1)m_{{b}}{m_{{c}}}^{2} \\
&&-5\,\mathit{I_{0}}(3,1,1)m_{{c}}{m_{{b}}}^{2}+15\,\mathit{I_{0}}(2,2,2)m_{{%
b}}{m_{{c}}}^{4}+20\,\mathit{I_{3}}(3,2,1)m_{{c}}{m_{{b}}}^{2}+5\,\mathit{%
I_{0}}^{{0,1}}(3,2,1)m_{{c}}{m_{{b}}}^{2} \\
&&-20\,\mathit{I_{0}}^{{0,1}}(3,2,1)m_{{b}}{m_{{c}}}^{2}-10\,\mathit{I_{0}}^{%
{0,1}}(3,1,2)m_{{b}}{m_{{c}}}^{2}+20\,\mathit{I_{0}}^{{0,1}}(3,1,2)m_{{c}}{%
m_{{b}}}^{2}-5\,\mathit{I_{0}}(3,2,1){m_{{c}}}^{3}{m_{{b}}}^{2} \\
&&+5\,\mathit{I_{0}}(3,2,1)m_{{b}}{m_{{c}}}^{4}-10\,\mathit{I_{0}}(3,2,1)m_{{%
c}}{m_{{b}}}^{4}-20\,\mathit{I_{3}}^{{0,1}}(3,2,2)m_{{c}}{m_{{b}}}^{2}+40\,%
\mathit{I_{3}}^{{0,1}}(3,2,2)m_{{b}}{m_{{c}}}^{2} \\
&&-60\,\mathit{I_{3}}(4,1,1)m_{{b}}{m_{{c}}}^{2}+30\,\mathit{I_{0}}(1,4,1)m_{%
{c}}{m_{{b}}}^{4}-30\,\mathit{I_{0}}(1,4,1){m_{{b}}}^{3}{m_{{c}}}^{2}-5\,%
\mathit{I_{0}}(3,2,2){m_{{b}}}^{3}{m_{{c}}}^{4} \\
&&+5\,\mathit{I_{0}}(3,2,2)m_{{b}}{m_{{c}}}^{6}-5\,\mathit{I_{0}}(3,2,2){m_{{%
c}}}^{5}{m_{{b}}}^{2}+5\,\mathit{I_{0}}(3,2,2){m_{{c}}}^{3}{m_{{b}}}%
^{4}-20\,m_{{b}}\mathit{I_{0}}^{{0,1}}(1,2,2) \\
&&+10\,m_{{b}}\mathit{I_{0}}^{{0,2}}(2,3,1)+10\,m_{{b}}\mathit{I_{0}}^{{0,1}%
}(1,3,1)-5\,m_{{b}}\mathit{I_{0}}^{{0,3}}(3,2,2)-15\,m_{{b}}\mathit{I_{0}}^{{%
0,1}}(2,2,1) \\
&&+30\,{m_{{b}}}^{3}\mathit{I_{0}}^{{0,1}}(1,4,1)+10\,m_{{b}}\mathit{I_{0}}^{%
{0,2}}(3,1,2)-15\,m_{{b}}{m_{{c}}}^{2}\mathit{I_{0}}^{{0,1}}(4,1,1)-25\,m_{{b%
}}\mathit{I_{0}}^{{0,1}}(2,1,2) \\
&&+15\,m_{{b}}\mathit{I_{0}}^{{0,2}}(2,2,2)+15\,m_{{b}}\mathit{I_{0}}^{{0,2}%
}(3,2,1)+40\,m_{{b}}\mathit{I_{3}}^{{0,1}}(2,3,1)+40\,m_{{b}}\mathit{I_{3}}%
(1,3,1) \\
&&+60\,\mathit{I_{3}}(4,1,1){m_{{c}}}^{3}+40\,\mathit{I_{3}}(3,2,1){m_{{c}}}%
^{3}+120\,\mathit{I_{3}}(1,4,1){m_{{b}}}^{3}+5\,\mathit{I_{0}}(1,2,2){m_{{b}}%
}^{3}+20\,\mathit{I_{3}}(3,1,2){m_{{c}}}^{3} \\
&&-15\,(\mathit{I_{0}}_{{0,1}})(3,2,1){m_{{b}}}^{3}-20\,\mathit{I_{3}}(3,2,1)%
{m_{{b}}}^{3}+40\,\mathit{I_{3}}(2,2,2){m_{{c}}}^{3}-20\,\mathit{I_{3}}%
(2,2,2){m_{{b}}}^{3} \\
&&-20\,\mathit{I_{3}}(2,2,1)m_{{c}}+40\,\mathit{I_{3}}(2,2,1)m_{{b}}+20\,m_{{%
b}}\mathit{I_{3}}(1,2,2)+20\,\mathit{I_{3}}^{{0,1}}(2,2,2)m_{{b}} \\
&&-15\,\mathit{I_{0}}(2,1,1)m_{{b}}+5\,\mathit{I_{0}}(2,1,1)m_{{c}}-40\,%
\mathit{I_{3}}(3,1,1)m_{{c}}-40\,\mathit{I_{3}}^{{0,1}}(2,2,2)m_{{c}} \\
&&+15\,\mathit{I_{0}}(1,2,1)m_{{b}}+40\,\mathit{I_{3}}(3,1,1)m_{{b}}-40\,%
\mathit{I_{3}}(2,3,1){m_{{b}}}^{3}-20\,\mathit{I_{3}}^{{0,1}}(3,1,2)m_{{c}}
\\
&&+40\,\mathit{I_{3}}^{{0,1}}(3,1,2)m_{{b}}-5\,\mathit{I_{0}}(1,2,1)m_{{c}%
}-40\,\mathit{I_{3}}^{{0,1}}(3,2,1)m_{{c}}+20\,\mathit{I_{3}}^{{0,1}%
}(3,2,1)m_{{b}} \\
&&+20\,\mathit{I_{3}}(2,1,2)m_{{b}}+5\,\mathit{I_{0}}^{{0,2}}(3,2,2){m_{{b}}}%
^{3}+20\,\mathit{I_{0}}(1,3,1){m_{{b}}}^{3}-20\,\mathit{I_{3}}(2,1,2)m_{{c}}
\\
&&-10\,\mathit{I_{0}}^{{0,1}}(3,1,1)m_{{b}}-5\,\mathit{I_{0}}^{{0,1}%
}(3,1,1)m_{{c}}-10\,\mathit{I_{0}}^{{0,1}}(2,3,1){m_{{b}}}^{3}+20\,\mathit{%
I_{3}}(3,2,2){m_{{c}}}^{5} \\
&&+20\,\mathit{I_{3}}^{{0,2}}(3,2,2)m_{{c}}-20\,\mathit{I_{3}}^{{0,2}%
}(3,2,2)m_{{b}}+15\,\mathit{I_{0}}(1,1,2)m_{{b}}-5\,\mathit{I_{0}}(1,1,2)m_{{%
c}} \\
&&-10\,\mathit{I_{0}}^{{0,1}}(2,2,2){m_{{b}}}^{3}
\end{eqnarray*}
\begin{eqnarray*}
C^{A'_{+}} &=&-20\mathit{I_{7}}^{{0,1}}(3,2,1)m_{{c}}-20\,\mathit{I_{4}}%
(3,1,1)m_{{c}}+10\mathit{I_{7}}^{{0,1}}(3,2,1)m_{{b}}-10\,\mathit{I_{4}}%
(2,2,2){m_{{b}}}^{3} \\
&&+20\,\mathit{I_{4}}(2,2,2){m_{{c}}}^{3}+60\,\mathit{I_{7}}(1,4,1){m_{{b}}}%
^{3}+20\,\mathit{I_{7}}(3,1,1)m_{{b}}-20\,\mathit{I_{7}}(3,1,1)m_{{c}} \\
&&+10\,\mathit{I_{4}}(3,2,2){m_{{c}}}^{5}-20\,\mathit{I_{4}}(3,2,1)m_{{b}}{%
m_{{c}}}^{2}-10\,\mathit{I_{4}}(3,1,2)m_{{b}}{m_{{c}}}^{2}-10\,\mathit{I_{7}}%
(3,1,2)m_{{b}}{m_{{c}}}^{2} \\
&&+20\,\mathit{I_{4}}(2,3,1)m_{{b}}{m_{{c}}}^{2}+10\,\mathit{I_{7}}(3,2,1){%
m_{{b}}}^{2}m_{{c}}-20\,\mathit{I_{7}}(3,2,1)m_{{b}}{m_{{c}}}^{2}-5\,\mathit{%
I_{0}}(3,2,2){m_{{c}}}^{3}{m_{{b}}}^{2} \\
&&-35\,\mathit{I_{1}}(3,2,1)m_{{b}}{m_{{c}}}^{2}-60\,\mathit{I_{4}}(1,4,1){%
m_{{b}}}^{2}m_{{c}}-30\,\mathit{I_{4}}(4,1,1)m_{{b}}{m_{{c}}}^{2}-10\,%
\mathit{I_{4}}(3,2,2)m_{{b}}{m_{{c}}}^{4} \\
&&-10\,\mathit{I_{4}}(3,2,2){m_{{c}}}^{3}{m_{{b}}}^{2}+10\,\mathit{I_{4}}%
(3,2,2){m_{{b}}}^{3}{m_{{c}}}^{2}-60\,\mathit{I_{7}}(1,4,1){m_{{b}}}^{2}m_{{c%
}}-10\,\mathit{I_{7}}(2,2,2){m_{{b}}}^{3} \\
&&+10\,\mathit{I_{4}}^{{0,1}}(3,2,2){m_{{b}}}^{3}-20\,\mathit{I_{4}}^{{0,1}%
}(3,2,2){m_{{c}}}^{3}-10\mathit{I_{0}}^{{0,1}}(3,2,2){m_{{c}}}^{3}+20\,%
\mathit{I_{7}}(2,2,2){m_{{c}}}^{3} \\
&&-10\mathit{I_{4}}^{{0,1}}(3,1,2)m_{{c}}+20\mathit{I_{4}}^{{0,1}}(3,1,2)m_{{%
b}}-10\,\mathit{I_{1}}(3,2,1){m_{{b}}}^{3}+10\,\mathit{I_{7}}(3,1,2){m_{{c}}}%
^{3} \\
&&+20\,\mathit{I_{4}}(2,2,1)m_{{b}}-10\,\mathit{I_{4}}(2,2,1)m_{{c}}+5%
\mathit{I_{1}}^{{0,1}}(3,2,2){m_{{b}}}^{3}+20\,\mathit{I_{1}}(2,1,2)m_{{b}}
\\
&&+10\,\mathit{I_{0}}(2,2,1)m_{{b}}+30\,\mathit{I_{1}}(1,4,1){m_{{b}}}^{3}+20%
\mathit{I_{7}}^{{0,1}}(3,1,2)m_{{b}}-10\mathit{I_{7}}^{{0,1}}(3,1,2)m_{{c}}
\\
&&-20\mathit{I_{7}}^{{0,1}}(2,2,2)m_{{c}}+5\,\mathit{I_{2}}(3,2,2){m_{{c}}}%
^{5}-10\mathit{I_{7}}^{{0,2}}(3,2,2)m_{{b}}+10\mathit{I_{7}}^{{0,2}%
}(3,2,2)m_{{c}} \\
&&+10\mathit{I_{4}}^{{0,2}}(3,2,2)m_{{c}}-10\mathit{I_{4}}^{{0,2}}(3,2,2)m_{{%
b}}+20\,m_{{b}}\mathit{I_{4}}(1,3,1)+20\,m_{{b}}\mathit{I_{1}}(1,2,2) \\
&&+20\,m_{{b}}\mathit{I_{4}}^{{0,1}}(2,3,1)+5\,m_{{c}}\mathit{I_{0}}%
(2,2,1)+30\,{m_{{c}}}^{3}\mathit{I_{1}}(2,2,2)+20\,{m_{{c}}}^{3}\mathit{I_{1}%
}(3,1,2) \\
&&+10\,m_{{b}}\mathit{I_{2}}(2,2,1)+15\,m_{{c}}\mathit{I_{2}}(2,1,2)+5\,{m_{{%
c}}}^{3}\mathit{I_{0}}(3,1,2)-20\,\mathit{I_{7}}(2,3,1){m_{{b}}}^{3}-20\,%
\mathit{I_{4}}(2,3,1){m_{{b}}}^{3} \\
&&+10\,{m_{{c}}}^{3}\mathit{I_{0}}(2,2,2)+10\,\mathit{I_{7}}(3,2,2){m_{{c}}}%
^{5}+10\,\mathit{I_{7}}(2,1,2)m_{{b}}-10\,\mathit{I_{7}}(2,1,2)m_{{c}}+10%
\mathit{I_{1}}^{{0,1}}(3,2,1)m_{{b}} \\
&&+10\,\mathit{I_{4}}(2,1,2)m_{{b}}-10\,\mathit{I_{4}}(2,1,2)m_{{c}}+5\,%
\mathit{I_{0}}(3,2,2){m_{{c}}}^{5}-5\mathit{I_{1}}^{{0,2}}(3,2,2)m_{{b}%
}+20\,m_{{b}}\mathit{I_{7}}(1,3,1) \\
&&-5\,m_{{c}}\mathit{I_{2}}(3,1,1)+25\,{m_{{c}}}^{3}\mathit{I_{0}}%
(3,2,1)-30\,m_{{c}}\mathit{I_{1}}^{{0,1}}(3,1,2)+15\,m_{{c}}\mathit{I_{1}}^{{%
0,2}}(3,2,2) \\
&&-15\,m_{{c}}\mathit{I_{1}}(3,1,1)+10\,m_{{c}}\mathit{I_{1}}(2,1,2)+30\,m_{{%
b}}\mathit{I_{1}}(2,2,1)-35\,m_{{c}}\mathit{I_{1}}^{{0,1}}(3,2,1)+45\,{m_{{c}%
}}^{3}\mathit{I_{1}}(4,1,1) \\
&&-30\,m_{{c}}\mathit{I_{1}}^{{0,1}}(2,2,2)+10\,m_{{c}}\mathit{I_{0}}%
(3,1,1)-10\,m_{{b}}\mathit{I_{2}}(1,3,1)-15\,m_{{c}}\mathit{I_{0}}^{{0,1}%
}(3,2,1) \\
&&-10\,m_{{c}}\mathit{I_{2}}^{{0,1}}(3,2,1)+10\,m_{{b}}\mathit{I_{1}}^{{0,1}%
}(2,3,1)+5\,m_{{c}}(\mathit{I_{0}}^{{0,2}})(3,2,2)+20\,\mathit{I_{7}}(3,2,1){%
m_{{c}}}^{3} \\
&&-10\,\mathit{I_{7}}(3,2,1){m_{{b}}}^{3}+15\,\mathit{I_{1}}(3,2,2){m_{{c}}}%
^{5}+10\mathit{I_{4}}^{{0,1}}(2,2,2)m_{{b}}+15\,{m_{{c}}}^{3}\mathit{I_{0}}%
(4,1,1) \\
&&+5\,m_{{c}}\mathit{I_{2}}^{{0,2}}(3,2,2)-10\,m_{{b}}\mathit{I_{0}}%
(1,3,1)-15\,m_{{c}}\mathit{I_{2}}^{{0,1}}(3,1,2)+15\,m_{{c}}\mathit{I_{0}}%
(2,1,2) \\
&&+20\,m_{{b}}\mathit{I_{7}}^{{0,1}}(2,3,1)-10\,m_{{b}}\mathit{I_{1}}%
(1,3,1)+15\,{m_{{c}}}^{3}\mathit{I_{2}}(4,1,1)+5\,{m_{{c}}}^{3}\mathit{I_{2}}%
(3,1,2) \\
&&-10\,m_{{c}}\mathit{I_{0}}^{{0,1}}(2,2,2)-10\,m_{{c}}\mathit{I_{2}}^{{0,1}%
}(2,2,2)-30\,m_{{c}}{m_{{b}}}^{2}\mathit{I_{0}}(1,4,1)+20\,m_{{b}}{m_{{c}}}%
^{2}\mathit{I_{2}}(2,3,1) \\
&&-30\,m_{{c}}{m_{{b}}}^{2}\mathit{I_{2}}(1,4,1)-15\,m_{{c}}\mathit{I_{0}}^{{%
0,1}}(3,1,2)-20\mathit{I_{4}}^{{0,1}}(2,2,2)m_{{c}}-20\mathit{I_{7}}^{{0,1}%
}(3,2,2){m_{{c}}}^{3} \\
&&+20\,\mathit{I_{4}}(3,1,1)m_{{b}}-5\,\mathit{I_{1}}(2,2,1)m_{{c}}+10\,{m_{{%
c}}}^{3}\mathit{I_{2}}(2,2,2)+30\,\mathit{I_{4}}(4,1,1){m_{{c}}}^{3}+10%
\mathit{I_{7}}^{{0,1}}(3,2,2){m_{{b}}}^{3} \\
&&+10\mathit{I_{1}}^{{0,1}}(3,1,2)m_{{b}}+10\mathit{I_{4}}^{{0,1}}(3,2,1)m_{{%
b}}-20\mathit{I_{4}}^{{0,1}}(3,2,1)m_{{c}}-30\mathit{I_{1}}^{{0,1}}(3,2,2){%
m_{{c}}}^{3} \\
&&-10\mathit{I_{2}}^{{0,1}}(3,2,2){m_{{c}}}^{3}+30\,\mathit{I_{7}}(4,1,1){m_{%
{c}}}^{3}+20\,\mathit{I_{4}}(3,2,1){m_{{c}}}^{3}+45\,\mathit{I_{1}}(3,2,1){%
m_{{c}}}^{3} \\
&&+10\mathit{I_{1}}^{{0,1}}(3,2,2)m_{{b}}{m_{{c}}}^{2}-5\,\mathit{I_{2}}%
(3,2,2){m_{{c}}}^{3}{m_{{b}}}^{2}-5\,\mathit{I_{1}}(3,2,2)m_{{b}}{m_{{c}}}%
^{4}+5\,\mathit{I_{1}}(3,2,2){m_{{b}}}^{3}{m_{{c}}}^{2} \\
&&+10\,\mathit{I_{4}}(3,1,2){m_{{c}}}^{3}+20\,\mathit{I_{7}}(2,2,1)m_{{b}%
}-10\,\mathit{I_{4}}(3,2,1){m_{{b}}}^{3}+20\,\mathit{I_{2}}(3,2,1){m_{{c}}}%
^{3}+60\,\mathit{I_{4}}(1,4,1){m_{{b}}}^{3}\\
&&-10\,\mathit{I_{7}}(2,2,1)m_{{c}}+10\,\mathit{I_{7}}^{{0,1}%
}(2,2,2)m_{{b}}-10\,\mathit{I_{1}}(2,3,1){m_{{b}}}^{3}+10\,m_{{b}}\mathit{%
I_{7}}(1,2,2) \\
&&+10\,m_{{b}}\mathit{I_{4}}(1,2,2)+20\,m_{{b}}{m_{{c}}}^{2}\mathit{I_{0}}%
(2,3,1)-90\,{m_{{b}}}^{2}m_{{c}}\mathit{I_{1}}(1,4,1)+50\,m_{{b}}{m_{{c}}}%
^{2}\mathit{I_{1}}(2,3,1) \\
&&+10\,\mathit{I_{7}}(3,2,2){m_{{b}}}^{3}{m_{{c}}}^{2}-15\,\mathit{I_{1}}%
(3,2,2){m_{{c}}}^{3}{m_{{b}}}^{2}+10\,\mathit{I_{1}}(2,2,2){m_{{b}}}^{2}m_{{c%
}}-15\mathit{I_{1}}^{{0,1}}(3,2,2){m_{{b}}}^{2}m_{{c}} \\
&&-30\,\mathit{I_{7}}(4,1,1)m_{{b}}{m_{{c}}}^{2}-5\,\mathit{I_{1}}(3,1,2){m_{%
{b}}}^{2}m_{{c}}-10\,\mathit{I_{4}}^{{0,1}}(3,2,2){m_{{b}}}^{2}m_{{c}}+10\,%
\mathit{I_{7}}(2,2,2){m_{{b}}}^{2}m_{{c}} \\
&&-10\,\mathit{I_{7}}(2,2,2)m_{{b}}{m_{{c}}}^{2}-15\,\mathit{I_{1}}(4,1,1)m_{%
{b}}{m_{{c}}}^{2}+20\,\mathit{I_{7}}^{{0,1}}(3,2,2)m_{{b}}{m_{{c}}}^{2}-10\,%
\mathit{I_{7}}^{{0,1}}(3,2,2){m_{{b}}}^{2}m_{{c}} \\
&&-5\,\mathit{I_{2}}^{{0,1}}(3,2,2){m_{{b}}}^{2}m_{{c}}-10\,\mathit{I_{1}}%
(3,1,2)m_{{b}}{m_{{c}}}^{2}+20\,\mathit{I_{7}}(2,3,1)m_{{b}}{m_{{c}}}%
^{2}-10\,\mathit{I_{7}}(3,2,2)m_{{b}}{m_{{c}}}^{4} \\
&&-10\,\mathit{I_{7}}(3,2,2){m_{{c}}}^{3}{m_{{b}}}^{2}-5\,\mathit{I_{0}}%
(3,2,1)m_{{b}}{m_{{c}}}^{2}+15\,\mathit{I_{0}}(3,2,1){m_{{b}}}^{2}m_{{c}%
}+20\,\mathit{I_{4}}^{{0,1}}(3,2,2)m_{{b}}{m_{{c}}}^{2} \\
&&+10\,\mathit{I_{4}}(2,2,2){m_{{b}}}^{2}m_{{c}}-10\,\mathit{I_{4}}(2,2,2)m_{%
{b}}{m_{{c}}}^{2}-5\,\mathit{I_{0}}^{{0,1}}(3,2,2){m_{{b}}}^{2}m_{{c}}+30\,%
\mathit{I_{1}}(3,2,1){m_{{b}}}^{2}m_{{c}} \\
&&-15\,\mathit{I_{2}}(3,2,1)m_{{b}}{m_{{c}}}^{2}+10\,\mathit{I_{2}}(3,2,1){%
m_{{b}}}^{2}m_{{c}}+10\,\mathit{I_{4}}(3,2,1){m_{{b}}}^{2}m_{{c}}
\end{eqnarray*}
\begin{eqnarray*}
C^{A'_{-}} &=&-10\,\mathit{I_{7}}(3,2,2){m_{{c}}}^{5}+10\,\mathit{I_{2}}^{{0,1}%
}(3,2,2){m_{{c}}}^{3}-10\,\mathit{I_{7}}^{{0,1}}(3,2,2){m_{{b}}}^{3}+20\,%
\mathit{I_{7}}^{{0,1}}(3,2,2){m_{{c}}}^{3} \\
&&-10\,\mathit{I_{1}}^{{0,1}}(2,2,2)m_{{b}}-10\,\mathit{I_{7}}(3,1,2){m_{{c}}%
}^{3}-5\,\mathit{I_{2}}(3,2,2){m_{{c}}}^{5}+10\,\mathit{I_{7}}^{{0,2}%
}(3,2,2)m_{{b}} \\
&&-10\,\mathit{I_{7}}^{{0,2}}(3,2,2)m_{{c}}+35\,\mathit{I_{1}}(3,2,1){m_{{c}}%
}^{3}+10\,\mathit{I_{0}}^{{0,1}}(3,2,2){m_{{c}}}^{3}+20\,\mathit{I_{7}}^{{0,1%
}}(2,2,2)m_{{c}} \\
&&-10\,\mathit{I_{7}}^{{0,1}}(2,2,2)m_{{b}}-20\,\mathit{I_{4}}^{{0,1}%
}(3,2,1)m_{{c}}+10\,\mathit{I_{4}}^{{0,1}}(3,2,1)m_{{b}}+10\,\mathit{I_{4}}^{%
{0,1}}(3,2,2){m_{{b}}}^{3} \\
&&-20\,\mathit{I_{4}}^{{0,1}}(3,2,2){m_{{c}}}^{3}+30\,\mathit{I_{4}}(4,1,1){%
m_{{c}}}^{3}-20\,\mathit{I_{7}}(3,2,1){m_{{c}}}^{3}-20\,\mathit{I_{4}}^{{0,1}%
}(2,2,2)m_{{c}} \\
&&+10\,\mathit{I_{4}}^{{0,1}}(2,2,2)m_{{b}}+10\,\mathit{I_{7}}(2,2,1)m_{{c}%
}-20\,\mathit{I_{7}}(2,2,1)m_{{b}}-10\,\mathit{I_{1}}^{{0,1}}(3,1,2)m_{{b}}
\\
&&-10\,\mathit{I_{4}}^{{0,2}}(3,2,2)m_{{b}}+10\,\mathit{I_{4}}^{{0,2}%
}(3,2,2)m_{{c}}-60\,\mathit{I_{7}}(1,4,1){m_{{b}}}^{3}+20\,\mathit{I_{4}}%
(2,2,2){m_{{c}}}^{3} \\
&&-5\,\mathit{I_{0}}(3,2,2){m_{{c}}}^{5}+10\,\mathit{I_{1}}(3,2,1){m_{{b}}}%
^{3}+20\,\mathit{I_{7}}^{{0,1}}(3,2,1)m_{{c}}-10\,\mathit{I_{7}}^{{0,1}%
}(3,2,1)m_{{b}} \\
&&-30\,\mathit{I_{1}}(1,4,1){m_{{b}}}^{3}+10\,\mathit{I_{4}}(3,2,2){m_{{c}}}%
^{5}-20\,\mathit{I_{1}}(2,1,2)m_{{b}}-10\,\mathit{I_{1}}^{{0,1}}(3,2,1)m_{{b}%
} \\
&&-20\,\mathit{I_{4}}(2,3,1){m_{{b}}}^{3}+10\,\mathit{I_{7}}(3,2,1){m_{{b}}}%
^{3}+10\,m_{{b}}\mathit{I_{4}}(1,2,2)+10\,\mathit{I_{1}}(3,1,2)m_{{b}}{m_{{c}%
}}^{2} \\
&&+20\,\mathit{I_{4}}^{{0,1}}(3,2,2)m_{{b}}{m_{{c}}}^{2}-10\,\mathit{I_{4}}^{%
{0,1}}(3,2,2)m_{{c}}{m_{{b}}}^{2}-30\,\mathit{I_{4}}(4,1,1)m_{{b}}{m_{{c}}}%
^{2} \\
&&-10\,\mathit{I_{7}}(3,2,1)m_{{c}}{m_{{b}}}^{2}+20\,\mathit{I_{7}}(3,2,1)m_{%
{b}}{m_{{c}}}^{2}-15\,\mathit{I_{0}}(3,2,1)m_{{c}}{m_{{b}}}^{2}+5\,\mathit{%
I_{0}}(3,2,1)m_{{b}}{m_{{c}}}^{2} \\
&&-10\,\mathit{I_{4}}(3,2,2){m_{{c}}}^{3}{m_{{b}}}^{2}-10\,\mathit{I_{4}}%
(3,2,2)m_{{b}}{m_{{c}}}^{4}+10\,\mathit{I_{4}}(3,2,2){m_{{b}}}^{3}{m_{{c}}}%
^{2}+20\,\mathit{I_{4}}(2,3,1)m_{{b}}{m_{{c}}}^{2} \\
&&-60\,\mathit{I_{4}}(1,4,1)m_{{c}}{m_{{b}}}^{2}-10\,\mathit{I_{7}}(2,2,2)m_{%
{c}}{m_{{b}}}^{2}+60\,\mathit{I_{7}}(1,4,1)m_{{c}}{m_{{b}}}^{2}+60\,\mathit{%
I_{4}}(1,4,1){m_{{b}}}^{3} \\
&&+10\,\mathit{I_{7}}(2,2,2){m_{{b}}}^{3}-20\,\mathit{I_{7}}(2,2,2){m_{{c}}}%
^{3}-20\,\mathit{I_{7}}(3,1,1)m_{{b}}+20\,\mathit{I_{7}}(3,1,1)m_{{c}} \\
&&+20\,\mathit{I_{7}}(2,3,1){m_{{b}}}^{3}-10\,\mathit{I_{4}}(2,2,2){m_{{b}}}%
^{3}+5\,\mathit{I_{1}}(2,2,1)m_{{c}}+10\,\mathit{I_{7}}(2,2,2)m_{{b}}{m_{{c}}%
}^{2} \\
&&-5\,\mathit{I_{1}}^{{0,1}}(3,2,2)m_{{c}}{m_{{b}}}^{2}-5\,\mathit{I_{1}}%
(3,2,2){m_{{c}}}^{3}{m_{{b}}}^{2}-10\,\mathit{I_{2}}(3,2,1)m_{{c}}{m_{{b}}}%
^{2}+15\,\mathit{I_{2}}(3,2,1)m_{{b}}{m_{{c}}}^{2} \\
&&+10\,\mathit{I_{4}}(3,2,1)m_{{c}}{m_{{b}}}^{2}-20\,\mathit{I_{4}}(3,2,1)m_{%
{b}}{m_{{c}}}^{2}+5\,\mathit{I_{1}}(3,2,2)m_{{b}}{m_{{c}}}^{4}-20\,\mathit{%
I_{7}}(2,3,1)m_{{b}}{m_{{c}}}^{2} \\
&&-5\,\mathit{I_{1}}(3,2,2){m_{{b}}}^{3}{m_{{c}}}^{2}-10\,\mathit{I_{4}}%
(2,2,2)m_{{b}}{m_{{c}}}^{2}+5\,\mathit{I_{0}}(3,2,2){m_{{c}}}^{3}{m_{{b}}}%
^{2}+15\,\mathit{I_{1}}(4,1,1)m_{{b}}{m_{{c}}}^{2} \\
&&-10\,\mathit{I_{4}}(3,1,2)m_{{b}}{m_{{c}}}^{2}-10\,\mathit{I_{1}}^{{0,1}%
}(3,2,2)m_{{b}}{m_{{c}}}^{2}+30\,\mathit{I_{7}}(4,1,1)m_{{b}}{m_{{c}}}%
^{2}+10\,\mathit{I_{7}}(3,2,2){m_{{c}}}^{3}{m_{{b}}}^{2} \\
&&+10\,\mathit{I_{7}}(3,2,2)m_{{b}}{m_{{c}}}^{4}-10\,\mathit{I_{7}}(3,2,2){%
m_{{b}}}^{3}{m_{{c}}}^{2}+5\,\mathit{I_{2}}^{{0,1}}(3,2,2)m_{{c}}{m_{{b}}}%
^{2}-20\,\mathit{I_{7}}^{{0,1}}(3,2,2)m_{{b}}{m_{{c}}}^{2} \\
&&+10\,\mathit{I_{7}}^{{0,1}}(3,2,2)m_{{c}}{m_{{b}}}^{2}+10\,\mathit{I_{7}}%
(3,1,2)m_{{b}}{m_{{c}}}^{2}+10\,\mathit{I_{4}}(2,2,2)m_{{c}}{m_{{b}}}%
^{2}+10\,\mathit{I_{7}}^{{0,1}}(3,1,2)m_{{c}} \\
&&-20\,\mathit{I_{7}}^{{0,1}}(3,1,2)m_{{b}}+20\,\mathit{I_{4}}(3,1,1)m_{{b}%
}-20\,\mathit{I_{4}}(3,1,1)m_{{c}}+10\,\mathit{I_{1}}(2,3,1){m_{{b}}}^{3} \\
&&-10\,\mathit{I_{4}}^{{0,1}}(3,1,2)m_{{c}}+20\,\mathit{I_{4}}^{{0,1}%
}(3,1,2)m_{{b}}-10\,\mathit{I_{4}}(2,2,1)m_{{c}}-5\,\mathit{I_{1}}^{{0,1}%
}(3,2,2){m_{{b}}}^{3} \\
&&-10\,\mathit{I_{0}}(2,2,1)m_{{b}}-10\,\mathit{I_{7}}(2,1,2)m_{{b}}+10\,%
\mathit{I_{1}}(2,2,2)m_{{b}}{m_{{c}}}^{2}-10\,\mathit{I_{1}}(2,2,2)m_{{c}}{%
m_{{b}}}^{2} \\
&&+10\,\mathit{I_{7}}(2,1,2)m_{{c}}-30\,\mathit{I_{7}}(4,1,1){m_{{c}}}%
^{3}-10\,\mathit{I_{1}}^{{0,1}}(3,2,2){m_{{c}}}^{3}+5\,\mathit{I_{1}}(3,2,2){%
m_{{c}}}^{5} \\
&&-20\,\mathit{I_{2}}(3,2,1){m_{{c}}}^{3}+20\,\mathit{I_{4}}(3,2,1){m_{{c}}}%
^{3}-10\,\mathit{I_{4}}(3,2,1){m_{{b}}}^{3}+10\,\mathit{I_{4}}(2,1,2)m_{{b}}
\\
&&-10\,\mathit{I_{4}}(2,1,2)m_{{c}}+5\,\mathit{I_{1}}^{{0,2}}(3,2,2)m_{{b}%
}+30\,m_{{b}}{m_{{c}}}^{2}\mathit{I_{1}}(2,3,1)-20\,m_{{b}}{m_{{c}}}^{2}%
\mathit{I_{2}}(2,3,1) \\
&&+30\,m_{{c}}{m_{{b}}}^{2}\mathit{I_{0}}(1,4,1)+30\,m_{{c}}{m_{{b}}}^{2}%
\mathit{I_{2}}(1,4,1)-30\,m_{{c}}{m_{{b}}}^{2}\mathit{I_{1}}(1,4,1)-20\,m_{{b%
}}{m_{{c}}}^{2}\mathit{I_{0}}(2,3,1) \\
&&-10\,m_{{b}}\mathit{I_{7}}(1,2,2)+10\,\mathit{I_{4}}(3,1,2){m_{{c}}}%
^{3}+5\,\mathit{I_{2}}(3,2,2){m_{{c}}}^{3}{m_{{b}}}^{2}+10\,\mathit{I_{1}}%
(3,2,1)m_{{c}}{m_{{b}}}^{2}\\&&-25\,\mathit{I_{1}}(3,2,1)m_{{b}}{m_{{c}}}^{2}+5\,\mathit{I_{0}}^{%
{0,1}}(3,2,2)m_{{c}}{m_{{b}}}^{2}+5\,\mathit{I_{1}}(3,1,2)m_{{c}}{m_{{b}}}%
^{2}-10\,{m_{{c}}}^{3}\mathit{I_{1}}(3,1,2) \\
&&-10\,m_{{c}}\mathit{I_{1}}^{{0,1}}(2,2,2)+10\,{m_{{c}}}^{3}\mathit{I_{1}}%
(2,2,2)+20\,m_{{b}}\mathit{I_{4}}(1,3,1)+20\,m_{{b}}\mathit{I_{1}}(2,2,1) \\
&&-15\,m_{{c}}\mathit{I_{1}}(3,1,1)+20\,m_{{b}}\mathit{I_{4}}^{{0,1}%
}(2,3,1)+15\,m_{{c}}\mathit{I_{0}}(2,2,1)+15\,{m_{{c}}}^{3}\mathit{I_{0}}%
(3,2,1) \\
&&-20\,m_{{c}}\mathit{I_{1}}^{{0,1}}(3,1,2)-5\,m_{{c}}\mathit{I_{1}}^{{0,1}%
}(3,2,1)+15\,{m_{{c}}}^{3}\mathit{I_{1}}(4,1,1)+5\,m_{{c}}\mathit{I_{1}}^{{%
0,2}}(3,2,2) \\
&&+10\,m_{{c}}\mathit{I_{2}}^{{0,1}}(3,2,1)-15\,m_{{c}}\mathit{I_{2}}%
(2,1,2)-15\,{m_{{c}}}^{3}\mathit{I_{2}}(4,1,1)+10\,m_{{b}}\mathit{I_{2}}%
(1,3,1) \\
&&+10\,m_{{c}}\mathit{I_{2}}^{{0,1}}(2,2,2)+10\,m_{{c}}\mathit{I_{0}}^{{0,1}%
}(2,2,2)-10\,{m_{{c}}}^{3}\mathit{I_{2}}(2,2,2)+10\,m_{{b}}\mathit{I_{0}}%
(1,3,1) \\
&&-30\,m_{{b}}\mathit{I_{1}}(1,3,1)-5\,m_{{c}}\mathit{I_{2}}^{{0,2}%
}(3,2,2)+15\,m_{{c}}\mathit{I_{2}}^{{0,1}}(3,1,2)+5\,m_{{c}}\mathit{I_{2}}%
(3,1,1) \\
&&-5\,{m_{{c}}}^{3}\mathit{I_{2}}(3,1,2)-10\,m_{{b}}\mathit{I_{1}}^{{0,1}%
}(2,3,1)-20\,m_{{b}}\mathit{I_{7}}^{{0,1}}(2,3,1)-5\,m_{{c}}\mathit{I_{0}}^{{%
0,2}}(3,2,2) \\
&&-15\,{m_{{c}}}^{3}\mathit{I_{0}}(4,1,1)-10\,{m_{{c}}}^{3}\mathit{I_{0}}%
(2,2,2)-15\,m_{{c}}\mathit{I_{0}}(2,1,2)-20\,m_{{b}}\mathit{I_{7}}(1,3,1) \\
&&+15\,m_{{c}}\mathit{I_{0}}^{{0,1}}(3,2,1)+15\,m_{{c}}\mathit{I_{0}}^{{0,1}%
}(3,1,2)-5\,{m_{{c}}}^{3}\mathit{I_{0}}(3,1,2)-10\,m_{{b}}\mathit{I_{2}}%
(2,2,1) \\
&&-10\,m_{{c}}\mathit{I_{0}}(3,1,1)-10\,m_{{b}}\mathit{I_{1}}(1,2,2)+20\,%
\mathit{I_{4}}(2,2,1)m_{{b}}
\end{eqnarray*}
\begin{eqnarray*}
C^{T'_{V}} &=&-10\,\mathit{I_{0}}^{{0,1}}(2,2,1)-5\mathit{I_{0}}^{{0,2}}(3,1,2)-10\,%
\mathit{I_{2}}(1,2,1)-10\,\mathit{I_{1}}(1,1,2)-5\,{m_{{b}}}^{2}\mathit{I_{0}%
}^{{0,2}}(3,2,2) \\
&&-10\,{m_{{b}}}^{2}\mathit{I_{1}}^{{0,2}}(3,2,2)-10\,{m_{{c}}}^{4}\mathit{%
I_{1}}(3,1,2)+20\,{m_{{b}}}^{2}\mathit{I_{1}}^{{0,1}}(2,2,2)+10\,{m_{{b}}}%
^{2}\mathit{I_{2}}(1,2,2) \\
&&+60\,\mathit{I_{0}}(2,1,1)-15\,{m_{{c}}}^{2}{m_{{b}}}^{2}\mathit{I_{0}}%
(4,1,1)+10\,{m_{{b}}}^{2}\mathit{I_{0}}^{{0,1}}(2,2,2)+20\,{m_{{b}}}^{3}m_{{c%
}}\mathit{I_{0}}(2,3,1) \\
&&-10\,{m_{{b}}}^{2}\mathit{I_{1}}(1,2,2)+10\,{m_{{c}}}^{2}\mathit{I_{2}}%
(3,1,1)-30\,{m_{{c}}}^{2}{m_{{b}}}^{2}\mathit{I_{1}}(4,1,1)+10\,{m_{{c}}}^{2}%
{m_{{b}}}^{2}\mathit{I_{2}}(3,2,1) \\
&&+40\,{m_{{b}}}^{3}m_{{c}}\mathit{I_{2}}(2,3,1)+20\mathit{I_{1}}^{{0,1}%
}(2,1,2)+5\,\mathit{I_{0}}^{{0,2}}(3,2,1)-15\,\mathit{I_{0}}(3,2,1){m_{{c}}}%
^{2}{m_{{b}}}^{2} \\
&&+10\,\mathit{I_{0}}^{{0,1}}(3,2,2){m_{{c}}}^{2}{m_{{b}}}^{2}+40\,\mathit{%
I_{2}}(2,2,1)m_{{b}}m_{{c}}-20\,\mathit{I_{1}}(3,2,1){m_{{c}}}^{2}{m_{{b}}}%
^{2}+10\,\mathit{I_{1}}(1,2,1) \\
&&-10\,\mathit{I_{1}}(2,2,2){m_{{b}}}^{4}+20\,\mathit{I_{1}}(2,2,1){m_{{c}}}%
^{2}+80\,\mathit{I_{1}}(2,2,1){m_{{b}}}^{2}+5\,\mathit{I_{0}}(3,2,1){m_{{c}}}%
^{4} \\
&&-10\,\mathit{I_{0}}(3,2,1){m_{{b}}}^{4}-5\,\mathit{I_{0}}(2,2,2){m_{{b}}}%
^{4}+20\mathit{I_{1}}^{{0,1}}(3,1,2){m_{{c}}}^{2}+5\mathit{I_{0}}^{{0,1}%
}(3,2,2){m_{{b}}}^{4} \\
&&+50\,\mathit{I_{2}}(2,2,1){m_{{b}}}^{2}+10\,\mathit{I_{1}}(3,2,1){m_{{c}}}%
^{4}-20\,\mathit{I_{1}}(3,2,1){m_{{b}}}^{4}-5\,\mathit{I_{0}}(3,2,2){m_{{c}}}%
^{4}{m_{{b}}}^{2} \\
&&-5\,\mathit{I_{0}}(3,1,1)m_{{b}}m_{{c}}+20\,\mathit{I_{0}}(2,2,1)m_{{b}}m_{%
{c}}+5\,\mathit{I_{0}}(3,1,2){m_{{c}}}^{2}{m_{{b}}}^{2}+20\mathit{I_{1}}^{{%
0,1}}(3,2,2){m_{{c}}}^{2}{m_{{b}}}^{2} \\
&&-10\,\mathit{I_{0}}(2,2,2){m_{{c}}}^{2}{m_{{b}}}^{2}-20\,\mathit{I_{1}}%
(2,2,2){m_{{c}}}^{2}{m_{{b}}}^{2}-10\,\mathit{I_{1}}(3,2,2){m_{{c}}}^{4}{m_{{%
b}}}^{2}+10\,\mathit{I_{1}}(3,2,2){m_{{c}}}^{2}{m_{{b}}}^{4} \\
&&-5\,\mathit{I_{0}}(3,2,1){m_{{b}}}^{3}m_{{c}}+5\,\mathit{I_{0}}(3,2,2){m_{{%
c}}}^{2}{m_{{b}}}^{4}+10\mathit{I_{1}}^{{0,1}}(3,1,1)-10\,\mathit{I_{0}}%
(3,1,1){m_{{b}}}^{2} \\
&&+20\mathit{I_{1}}^{{0,1}}(3,1,2){m_{{b}}}^{2}+10\,\mathit{I_{0}}^{{0,1}%
}(3,1,2){m_{{c}}}^{2}-10\,\mathit{I_{0}}(2,1,2){m_{{c}}}^{2}-20\,\mathit{%
I_{0}}(2,1,2){m_{{b}}}^{2} \\
&&-20\,\mathit{I_{1}}(2,1,2){m_{{c}}}^{2}-30\,\mathit{I_{1}}(2,1,2){m_{{b}}}%
^{2}-20\mathit{I_{1}}^{{0,1}}(3,2,1){m_{{c}}}^{2}+10\mathit{I_{1}}^{{0,1}%
}(3,2,1){m_{{b}}}^{2} \\
&&+10\,\mathit{I_{0}}(2,2,1){m_{{c}}}^{2}+50\,\mathit{I_{0}}(2,2,1){m_{{b}}}%
^{2}-5\,\mathit{I_{0}}(3,1,2){m_{{c}}}^{4}+15\mathit{I_{0}}^{{0,1}}(3,1,2){%
m_{{b}}}^{2} \\
&&-10\mathit{I_{0}}^{{0,1}}(3,2,1){m_{{c}}}^{2}+5\mathit{I_{0}}^{{0,1}%
}(3,2,1){m_{{b}}}^{2}+10\mathit{I_{1}}^{{0,1}}(3,2,2){m_{{b}}}^{4}-10\,%
\mathit{I_{0}}(3,1,1){m_{{c}}}^{2} \\
&&+50\,\mathit{I_{2}}(2,1,1)-10\mathit{I_{1}}^{{0,2}}(3,1,2)+10\mathit{I_{1}}%
^{{0,2}}(3,2,1)+90\,\mathit{I_{1}}(2,1,1)+10\,\mathit{I_{0}}^{{0,1}}(2,1,2)
\\
&&+10\,\mathit{I_{2}}(1,1,2)-20\mathit{I_{1}}^{{0,1}}(2,2,1)
\end{eqnarray*}
\begin{eqnarray*}
C^{T'_{0}} &=&-15\,{m_{{b}}}^{2}\mathit{I_{0}}^{{0,2}}(2,2,2)-5\mathit{I_{0}}%
^{{0,3}}(3,2,1)+5\,{m_{{b}}}^{2}\mathit{I_{0}}^{{0,3}}(3,2,2)-20\mathit{I_{0}%
}^{{0,1}}(3,1,2){m_{{b}}}^{3}m_{{c}} \\
&&-50\,\mathit{I_{0}}(1,2,1)m_{{b}}m_{{c}}+20\mathit{I_{0}}^{{0,1}}(2,2,1)m_{%
{b}}m_{{c}}+10\,\mathit{I_{0}}(2,1,2)m_{{b}}{m_{{c}}}^{3}+15\,\mathit{I_{0}}%
(2,1,2){m_{{b}}}^{3}m_{{c}} \\
&&-15\,\mathit{I_{0}}(2,1,2){m_{{c}}}^{2}{m_{{b}}}^{2}+10\,\mathit{I_{0}}%
(1,2,2){m_{{b}}}^{3}m_{{c}}+5\,\mathit{I_{0}}(3,1,2)m_{{b}}{m_{{c}}}^{5}+10%
\mathit{I_{0}}^{{0,1}}(3,1,1)m_{{b}}m_{{c}} \\
&&-5\,\mathit{I_{0}}(3,2,2){m_{{c}}}^{6}{m_{{b}}}^{2}-5\,\mathit{I_{0}}%
(3,2,2){m_{{b}}}^{5}{m_{{c}}}^{3}+5\,\mathit{I_{0}}(3,2,2){m_{{b}}}^{3}{m_{{c%
}}}^{5}+5\,\mathit{I_{0}}(3,2,2){m_{{c}}}^{4}{m_{{b}}}^{4} \\
&&+30\,\mathit{I_{0}}(1,3,1){m_{{c}}}^{2}{m_{{b}}}^{2}-70\,\mathit{I_{0}}%
(1,3,1){m_{{b}}}^{3}m_{{c}}-30\,\mathit{I_{0}}(2,1,1)m_{{b}}m_{{c}}-10%
\mathit{I_{0}}^{{0,1}}(2,1,2)m_{{b}}m_{{c}} \\
&&+30\,\mathit{I_{0}}(1,4,1){m_{{c}}}^{2}{m_{{b}}}^{4}-30\,\mathit{I_{0}}%
(1,4,1){m_{{b}}}^{5}m_{{c}}+15\mathit{I_{0}}^{{0,1}}(3,1,2){m_{{c}}}^{2}{m_{{%
b}}}^{2}-10\mathit{I_{0}}^{{0,1}}(3,1,2)m_{{b}}{m_{{c}}}^{3} \\
&&-10\,\mathit{I_{0}}(2,3,1){m_{{b}}}^{3}{m_{{c}}}^{3}-15\,\mathit{I_{0}}%
(4,1,1){m_{{c}}}^{4}{m_{{b}}}^{2}+15\,\mathit{I_{0}}(4,1,1){m_{{b}}}^{3}{m_{{%
c}}}^{3}-15\mathit{I_{0}}^{{0,2}}(3,2,2){m_{{c}}}^{2}{m_{{b}}}^{2} \\
&&+5\mathit{I_{0}}^{{0,2}}(3,2,2){m_{{b}}}^{3}m_{{c}}+15\mathit{I_{0}}^{{0,1}%
}(3,2,2){m_{{c}}}^{4}{m_{{b}}}^{2}-10\mathit{I_{0}}^{{0,1}}(3,2,2){m_{{b}}}%
^{3}{m_{{c}}}^{3}-5\mathit{I_{0}}^{{0,1}}(3,2,2){m_{{b}}}^{5}m_{{c}} \\
&&+10\,\mathit{I_{0}}(3,2,1){m_{{b}}}^{3}{m_{{c}}}^{3}-5\,\mathit{I_{0}}%
(3,2,1)m_{{b}}{m_{{c}}}^{5}-10\,\mathit{I_{0}}(3,2,1){m_{{c}}}^{4}{m_{{b}}}%
^{2}+10\,\mathit{I_{0}}(1,1,2)m_{{b}}m_{{c}} \\
&&+10\,\mathit{I_{0}}(2,3,1){m_{{b}}}^{5}m_{{c}}+5\mathit{I_{0}}^{{0,2}%
}(3,1,2)m_{{b}}m_{{c}}-5\,\mathit{I_{0}}(3,1,1){m_{{c}}}^{2}{m_{{b}}}^{2}+5\,%
\mathit{I_{0}}(3,1,1){m_{{b}}}^{3}m_{{c}} \\
&&-20\,\mathit{I_{0}}(1,2,2){m_{{c}}}^{2}{m_{{b}}}^{2}-5\mathit{I_{0}}^{{0,2}%
}(3,2,1)m_{{b}}m_{{c}}-5\,\mathit{I_{0}}(2,2,1){m_{{b}}}^{3}m_{{c}}-20\,%
\mathit{I_{0}}(2,2,1)m_{{b}}{m_{{c}}}^{3} \\
&&-10\,\mathit{I_{0}}(3,2,1){m_{{c}}}^{2}{m_{{b}}}^{4}+10\,\mathit{I_{0}}%
(3,2,1){m_{{b}}}^{5}m_{{c}}-15\,\mathit{I_{0}}(2,2,2){m_{{c}}}^{4}{m_{{b}}}%
^{2}+10\,\mathit{I_{0}}(2,2,2){m_{{b}}}^{3}{m_{{c}}}^{3} \\
&&+30\mathit{I_{0}}^{{0,1}}(2,2,2){m_{{c}}}^{2}{m_{{b}}}^{2}-10\mathit{I_{0}}%
^{{0,1}}(2,2,2){m_{{b}}}^{3}m_{{c}}+15\mathit{I_{0}}^{{0,1}}(3,2,1){m_{{c}}}%
^{2}{m_{{b}}}^{2}+10\mathit{I_{0}}^{{0,1}}(3,2,1)m_{{b}}{m_{{c}}}^{3} \\
&&-30\,{m_{{b}}}^{4}\mathit{I_{0}}^{{0,1}}(1,4,1)+10\,{m_{{b}}}^{3}m_{{c}}%
\mathit{I_{0}}^{{0,1}}(2,3,1)+20\mathit{I_{0}}^{{0,1}}(1,1,2)+15\,{m_{{c}}}%
^{2}{m_{{b}}}^{2}\mathit{I_{0}}^{{0,1}}(4,1,1) \\
&&-20\,\mathit{I_{0}}(1,1,2){m_{{c}}}^{2}-15\mathit{I_{0}}^{{0,1}}(3,2,1){m_{%
{c}}}^{4}+10\mathit{I_{0}}^{{0,1}}(2,2,2){m_{{b}}}^{4}+5\,\mathit{I_{0}}%
(3,2,1){m_{{c}}}^{6} \\
&&-5\mathit{I_{0}}^{{0,2}}(3,2,2){m_{{b}}}^{4}-15\,\mathit{I_{0}}(1,1,2){m_{{%
b}}}^{2}-5\mathit{I_{0}}^{{0,2}}(3,2,1){m_{{b}}}^{2}+20\,{m_{{b}}}^{2}%
\mathit{I_{0}}^{{0,1}}(1,2,2) \\
&&+15\,\mathit{I_{0}}(2,2,1){m_{{c}}}^{4}+10\,\mathit{I_{0}}(2,2,1){m_{{b}}}%
^{4}+10\mathit{I_{0}}^{{0,1}}(3,2,1){m_{{b}}}^{4}+15\mathit{I_{0}}^{{0,2}%
}(3,2,1){m_{{c}}}^{2} \\
&&+15\mathit{I_{0}}^{{0,1}}(3,1,2){m_{{c}}}^{4}+20\,\mathit{I_{0}}(2,1,1){m_{%
{b}}}^{2}+30\mathit{I_{0}}^{{0,1}}(2,1,2){m_{{c}}}^{2}+30\mathit{I_{0}}^{{0,1%
}}(2,1,2){m_{{b}}}^{2} \\
&&+15\,\mathit{I_{0}}(2,1,1){m_{{c}}}^{2}-15\,\mathit{I_{0}}(2,1,2){m_{{c}}}%
^{4}-5\mathit{I_{0}}^{{0,1}}(2,2,1){m_{{b}}}^{2}-30\mathit{I_{0}}^{{0,1}%
}(2,2,1){m_{{c}}}^{2} \\
&&+10\mathit{I_{0}}^{{0,1}}(3,1,1){m_{{b}}}^{2}-5\,\mathit{I_{0}}(1,2,2){m_{{%
b}}}^{4}-15\mathit{I_{0}}^{{0,2}}(3,1,2){m_{{b}}}^{2}+20\,\mathit{I_{0}}%
(1,2,1){m_{{c}}}^{2} \\
&&+10\,\mathit{I_{0}}(1,2,1){m_{{b}}}^{2}-15\mathit{I_{0}}^{{0,2}}(3,1,2){m_{%
{c}}}^{2}-5\,\mathit{I_{0}}(3,1,2){m_{{c}}}^{6}-30\,{m_{{b}}}^{2}\mathit{%
I_{0}}^{{0,1}}(1,3,1) \\
&&+5\mathit{I_{0}}^{{0,3}}(3,1,2)+15\,\mathit{I_{0}}(1,1,1)-20\mathit{I_{0}}%
^{{0,1}}(1,2,1)-15\mathit{I_{0}}^{{0,1}}(2,1,1) \\
&&-15\mathit{I_{0}}^{{0,2}}(2,1,2)+15\mathit{I_{0}}^{{0,2}}(2,2,1)
\end{eqnarray*}
\begin{eqnarray*}
C^{T'_{-}} &=&5/2\,{m_{{b}}}^{2}\mathit{I_{0}}^{{0,2}}(3,2,2)-5\,\mathit{I_{0}}%
(2,1,1)-5/2\,\mathit{I_{0}}^{{0,2}}(3,2,1)-5\,\mathit{I_{0}}(1,2,1) \\
&&-5\,\mathit{I_{1}}^{{0,2}}(3,1,2)-20\,{m_{{b}}}^{3}m_{{c}}\mathit{I_{1}}%
(2,3,1)-15\,{m_{{c}}}^{2}{m_{{b}}}^{2}\mathit{I_{1}}(4,1,1)+10\,\mathit{I_{1}%
}^{{0,1}}(2,1,2) \\
&&+15/2\,{m_{{c}}}^{2}{m_{{b}}}^{2}\mathit{I_{0}}(4,1,1)+10\,{m_{{b}}}^{3}m_{%
{c}}\mathit{I_{0}}(2,3,1)-15/2\,{m_{{c}}}^{2}{m_{{b}}}^{2}\mathit{I_{0}}%
(3,2,1)-15\,\mathit{I_{1}}(2,1,1) \\
&&+5\,\mathit{I_{0}}(3,1,1){m_{{b}}}^{2}+5\,\mathit{I_{0}}^{{0,1}%
}(2,2,1)+5/2\,\mathit{I_{0}}^{{0,2}}(3,1,2)-10\,{m_{{c}}}^{2}\mathit{I_{1}}^{%
{0,1}}(3,2,1) \\
&&-5\,{m_{{b}}}^{2}\mathit{I_{0}}^{{0,1}}(2,2,2)-5\,{m_{{b}}}^{2}\mathit{%
I_{1}}^{{0,2}}(3,2,2)+5\,{m_{{b}}}^{2}\mathit{I_{0}}(1,2,2)-5\,{m_{{c}}}^{4}%
\mathit{I_{1}}(3,1,2) \\
&&-10\,{m_{{c}}}^{2}\mathit{I_{0}}(3,1,1)+30\,{m_{{b}}}^{4}\mathit{I_{1}}%
(1,4,1)+10\,{m_{{b}}}^{2}\mathit{I_{1}}^{{0,1}}(2,2,2)-10\,\mathit{I_{1}}^{{%
0,1}}(2,2,1) \\
&&-15\,{m_{{b}}}^{4}\mathit{I_{0}}(1,4,1)-15\,{m_{{b}}}^{2}\mathit{I_{0}}%
(1,3,1)+30\,{m_{{b}}}^{2}\mathit{I_{1}}(1,3,1)+5\,\mathit{I_{0}}(2,2,2){m_{{c%
}}}^{2}{m_{{b}}}^{2} \\
&&+10\,\mathit{I_{1}}(2,1,2)m_{{b}}m_{{c}}+10\,\mathit{I_{0}}(2,2,1)m_{{b}%
}m_{{c}}-30\,\mathit{I_{1}}(2,2,1)m_{{b}}m_{{c}}-5\,\mathit{I_{0}}^{{0,1}%
}(3,2,2){m_{{c}}}^{2}{m_{{b}}}^{2} \\
&&+5\,\mathit{I_{1}}(3,1,2){m_{{c}}}^{2}{m_{{b}}}^{2}-5\,\mathit{I_{1}}^{{0,1%
}}(3,1,1)+5\,\mathit{I_{1}}^{{0,2}}(3,2,1)+5/2\,\mathit{I_{0}}(2,2,2){m_{{b}}%
}^{4} \\
&&+15\,\mathit{I_{1}}(2,1,2){m_{{b}}}^{2}-5\,\mathit{I_{0}}(2,2,1){m_{{c}}}%
^{2}+5\,\mathit{I_{1}}(3,1,1){m_{{b}}}^{2}+10\,\mathit{I_{1}}(2,2,1){m_{{c}}}%
^{2} \\
&&-10\,\mathit{I_{1}}(2,1,2){m_{{c}}}^{2}-5/2\,\mathit{I_{0}}^{{0,1}}(3,2,2){%
m_{{b}}}^{4}+10\,\mathit{I_{1}}^{{0,1}}(3,1,2){m_{{c}}}^{2}+15\,\mathit{I_{1}%
}^{{0,1}}(3,1,2){m_{{b}}}^{2} \\
&&-5\,\mathit{I_{1}}(3,1,1){m_{{c}}}^{2}+5/2\,\mathit{I_{0}}(3,1,2){m_{{c}}}%
^{4}+5\,\mathit{I_{1}}^{{0,1}}(3,2,2){m_{{b}}}^{4}+10\,\mathit{I_{0}}(2,1,2){%
m_{{b}}}^{2} \\
&&+5\,\mathit{I_{0}}(2,1,2){m_{{c}}}^{2}-5/2\,\mathit{I_{0}}(3,2,1){m_{{c}}}%
^{4}-5\,\mathit{I_{1}}(3,2,1){m_{{b}}}^{3}m_{{c}}+5/2\,\mathit{I_{0}}(3,2,2){%
m_{{c}}}^{4}{m_{{b}}}^{2} \\
&&-5/2\,\mathit{I_{0}}(3,2,2){m_{{c}}}^{2}{m_{{b}}}^{4}+10\,\mathit{I_{1}}^{{%
0,1}}(3,2,2){m_{{c}}}^{2}{m_{{b}}}^{2}-5\,\mathit{I_{1}}(3,1,2){m_{{b}}}%
^{3}m_{{c}}+10\,\mathit{I_{1}}(2,2,2){m_{{b}}}^{3}m_{{c}} \\
&&-10\,\mathit{I_{1}}(2,2,2){m_{{c}}}^{2}{m_{{b}}}^{2}-5\,\mathit{I_{1}}%
(3,2,2){m_{{c}}}^{4}{m_{{b}}}^{2}+5\,\mathit{I_{1}}(3,2,2){m_{{c}}}^{2}{m_{{b%
}}}^{4}-5/2\,\mathit{I_{0}}(3,1,2){m_{{c}}}^{2}{m_{{b}}}^{2} \\
&&+5/2\,\mathit{I_{0}}(3,2,1){m_{{b}}}^{3}m_{{c}}-10\,\mathit{I_{1}}(3,2,1){%
m_{{c}}}^{2}{m_{{b}}}^{2}+5\,\mathit{I_{0}}(3,2,1){m_{{b}}}^{4}+5\,\mathit{%
I_{1}}(3,2,1){m_{{c}}}^{4} \\
&&-5\,\mathit{I_{1}}(3,2,1){m_{{b}}}^{4}-5\,\mathit{I_{0}}^{{0,1}}(3,1,2){m_{%
{c}}}^{2}-15/2\,\mathit{I_{0}}^{{0,1}}(3,1,2){m_{{b}}}^{2}+5\,\mathit{I_{0}}%
^{{0,1}}(3,2,1){m_{{c}}}^{2} \\
&&-5/2\,\mathit{I_{0}}^{{0,1}}(3,2,1){m_{{b}}}^{2}+5/2\,\mathit{I_{0}}%
(3,1,1)m_{{b}}m_{{c}}-5\,\mathit{I_{0}}^{{0,1}}(2,1,2)+5\,\mathit{I_{0}}%
(1,1,2)
\end{eqnarray*}
where \bea \hat{I}_n^{[i,j]} (a,b,c) = \ga M_1^2 \dr^i \ga M_2^2
\dr^j \frac{d^i}{d\ga M_1^2 \dr^i} \frac{d^j}{d\ga M_2^2 \dr^j}
\left[\ga M_1^2 \dr^i \ga M_2^2 \dr^j \hat{I}_n(a,b,c) \right]~.
\nnb \eea
\end{document}